\documentclass{jfm}

\usepackage{subfigure}
\usepackage{graphicx}
\usepackage{epstopdf, epsfig}
\usepackage{graphicx}
\usepackage{epsfig}
\usepackage{bm}
\usepackage{amsmath}
\usepackage{amssymb}
\usepackage{wasysym}
\usepackage{epstopdf}
\usepackage{color}
\usepackage{xcolor}
\usepackage{epstopdf}
\usepackage{upgreek}
\usepackage{natbib}
\usepackage{arydshln}
\usepackage[T1]{fontenc}

\usepackage{natbib}

\usepackage{psfrag}
\usepackage{color}
\usepackage{graphicx,float,amssymb}

\usepackage{color}
\usepackage{graphicx}
\usepackage{natbib}
\usepackage{bm}
\usepackage{amsmath}

\usepackage[colorlinks=false, bookmarks=true]{hyperref}

\usepackage{graphicx}
\usepackage{epsfig}
\usepackage{bm}
\usepackage{amsmath}
\usepackage{amssymb}
\usepackage{wasysym}
\usepackage{epstopdf}
\usepackage{color}
\usepackage{xcolor}
\usepackage{epstopdf}
\usepackage{upgreek}
\usepackage{natbib}
\usepackage{arydshln}
\usepackage[T1]{fontenc}

\shortauthor{E. Alm\'eras, V. Mathai, D. Lohse and C. Sun}

\title{Experimental investigation of the turbulence induced by a bubble swarm rising within an incident turbulence}

\author{Elise Alm\'eras\aff{1},
	Varghese Mathai\aff{1},
	Detlef Lohse\aff{1,2},
	and Chao Sun\aff{3,1} \corresp{\email{chaosun@tsinghua.edu.cn}}}

\affiliation{\aff{1} Physics of Fluids Group, Faculty of Science and Technology, J.M. Burgers Center for Fluid Dynamics,
	University of Twente, P.O. Box 217, 7500 AE Enschede, The Netherlands
	
	\aff{2} Max Planck Institute for Dynamics and Self-Organization, 37077 G\"ottingen, Germany.
	\aff{3} Center for Combustion Energy and Department of Thermal Engineering, Tsinghua University, Beijing 100084, China}

\begin{document}
	
	\maketitle

	\begin{abstract} 
		This work reports an experimental characterisation of the flow properties in a homogeneous bubble swarm rising at high-Reynolds-numbers within a homogeneous and isotropic turbulent flow. Both the gas volume fraction $\alpha$ and the velocity fluctuations $u'_0$ of the carrier flow before bubble injection are varied, respectively, in the ranges $0 \% <\alpha<0.93 \%$ and $2.3~\text{cm/s}<u'_0 <5.5$ cm/s. The so-called bubblance parameter~($b=\frac{V_r^2 \alpha}{u'^2_0}$, where $V_r$ is the bubble rise velocity) is used to compare the ratio of the kinetic energy generated by the bubbles to the one produced by the incident turbulence, and is varied from 0 to 1.3. Conditional measurements of the velocity field downstream of the bubbles {\color{black}in the vertical direction} allow us to disentangle three regions that have specific statistical properties, namely the primary wake, the secondary wake and the far field. While the fluctuations in the primary wake are similar to the one of a single bubble rising in a liquid at rest, the statistics of the velocity fluctuations in the far field follow Gaussian distribution, similar to the one produced by the homogenous and isotropic turbulence at the largest scales. In the secondary wake region, the conditional probability density function (pdf) of the velocity fluctuations is asymmetric and shows an exponential tail for the positive fluctuations and a Gaussian one for the negative fluctuations. The overall agitation thus results from the combination of these three contributions and depends mainly on the bubblance parameter. For $0<b<0.7$, the overall velocity fluctuations {\color{black}in the vertical direction} evolve as $b^{0.4}$ and are mostly driven by the far field agitation, whereas the fluctuations increase as $b^{1.3}$ for larger values of the bubblance parameter ($b>0.7$), in which the significant contributions come both from the secondary wake and the far field. Thus, the bubblance parameter is a suitable parameter to characterise the evolution of liquid agitation in bubbly turbulent flows.
	\end{abstract}
	
	\begin{keywords} bubbly flow, incident turbulence, wakes 
	\end{keywords}

\section{Introduction}

Numerous industrial processes, such as the Fischer-Tropsch and reaction catalysis make use of the agitation induced by bubbles to enhance mixing and mass transfer. Thanks to their buoyancy, bubbles rise at a higher velocity than that of the liquid phase, thus inducing perturbation therein. However, the turbulence induced by bubbles, commonly called pseudo-turbulence, differs a lot from homogeneous-isotropic-turbulence or wall-bounded turbulence. In the last two decades, several studies have been conducted on this subject, both experimentally~(\cite{risso2002velocity,riboux2010experimental,mendez2013power,roghair2011energy,martinez2010bubble,lance1991turbulence}) and numerically~(\cite{balachandar2010turbulent,riboux2013model,van2008numerical,darmana2005detailed,roghair2011energy,mazzitelli2004lagrangian}, leading to a clearer understanding of the bubble-induced liquid agitation. The liquid phase fluctuations due to pseudo-turbulence result mainly from two contributions, one from the bubble wakes themselves, and the other one from the non-linear interactions between the bubble wakes. However, these two contributions do not have the same role. While the fluctuations contained in the wake are mostly upward and cause a positive skewness of the pdf of the velocity fluctuations in the vertical direction, most of the liquid agitation comes from the interaction between the wakes, which result in exponential tails of the velocity pdfs. A consequence of this interaction between the wakes is that the velocity fluctuations have a non-linear dependence on the gas volume fraction, as: $V_r \alpha^{0.5}$, where $V_r$ is the relative rising velocity of the bubbles with respect to the carrier fluid, and $\alpha$ is the gas volume fraction. Another feature of pseudo-turbulence is that it is inherently anisotropic, meaning that the vertical fluctuations produced by the bubbles are stronger as compared to the horizontal ones. 

While pseudo-turbulence has been adequately characterized, in most natural and industrial settings, bubble-induced agitation often occurs in the presence of surrounding turbulence. This leads to a more complex agitation, since it results from the coupling of these two sources of turbulence.
A very different resulting agitation can be observed in such a bubble swarm rising in a turbulent background flow, depending on the nature of the surrounding turbulence, on the ratio of the energies produced by the two sources, and on their characteristic length- and time-scales. In this work, we focus on the agitation produced by a bubble swarm rising within a nearly homogenous and isotropic turbulent flow. One parameter, which is commonly used to characterise such a turbulent flow, is the so-called \textit{bubblance} parameter $b$. It is defined to be proportional to $\frac{V_r^2 \alpha}{u'^2_0}$, where $u'^2_0$ is the variance of the velocity fluctuations produced by the external turbulence in the absence of bubbles~(\cite{rensen2005effect,prakash2016energy,lance1991turbulence}). By definition, $b=0$ corresponds to single-phase flow, whereas $b=\infty$ corresponds to a bubble swarm rising in a quiescent liquid~(pseudo-turbulence). A wide range of bubblance parameters are thus possible between these two extreme limits, where both pseudo-turbulence and homogeneous and isotropic turbulence are coupled. However, few studies exist in this domain apart from the spectral analyses of \cite{lance1991turbulence} \& \cite{prakash2016energy}. 

Another important feature of turbulent flows in general is the wide range of length and time scales. The energy contained at these scales can be captured by the spectrum of velocity fluctuations. For homogeneous-isotropic-turblence, the spectrum of velocity fluctuations shows the classical $-5/3$ scaling for the inertial sub-range in both wavenumber and frequency domains. However, for pseudo-turbulence the spectrum of the velocity fluctuations displays a signature -3 scaling~(\cite{lance1991turbulence}). This originates when the energy put in by the bubbles into the fluid by the wake is directly dissipated. Consequently, the $-3$ scaling is known to start at a wavenumber $\lambda^{-1}=\frac{C_d}{d}$, where $C_d$ is the drag coefficient of the rising bubble, and $d$ is its diameter~(\cite{riboux2010experimental,roghair2011energy}).

The purpose of the present paper is to study the properties of the liquid agitation in turbulent bubbly flows, in the domain where both homogeneous isotropic turbulence and pseudo-turbulence play a role, and to disentangle the different sources of agitation. This leads us to explore the changes to the wake behind the bubble when the bubblance parameter is varied. {\color{black} A similar range of bubblance parameter variation was studied by~\cite{prakash2016energy}. However, they focussed only on the normalized velocity PDFs and energy spectra. Here, in addition to these analyses, we present a conditional analysis of the statistical properties. This allows us to distinguish three regions in the flow (the primary wake, the secondary wake, and the far field), whose the physical properties are fundamentally different.}

The paper is organised as follows. First, the setup and the measurement techniques are presented in section~\ref{sec:setup}. Then, in section~\ref{sec:op_cond} we discuss the operating conditions in order to characterise the agitation produced by the incident turbulence in the absence of bubbles and the properties of the bubbles. The dynamics of the liquid phase is then explored by considering two approaches. The first approach consists in investigating the overall properties of the liquid phase~(section \ref{sec:dyn_liq_Ncond}), and the second one, the conditional statistics of the flow properties~(section \ref{sec:cond_properties}). A discussion, in section~\ref{sec:discussion}, is carried out to link the overall and conditional statistics before concluding with the main findings and future recommendations~(section \ref{sec:concl}).

\section{Experimental set-up and instrumentation}
\label{sec:setup}

\begin{figure}
	\centering	
	\includegraphics[scale=.32]{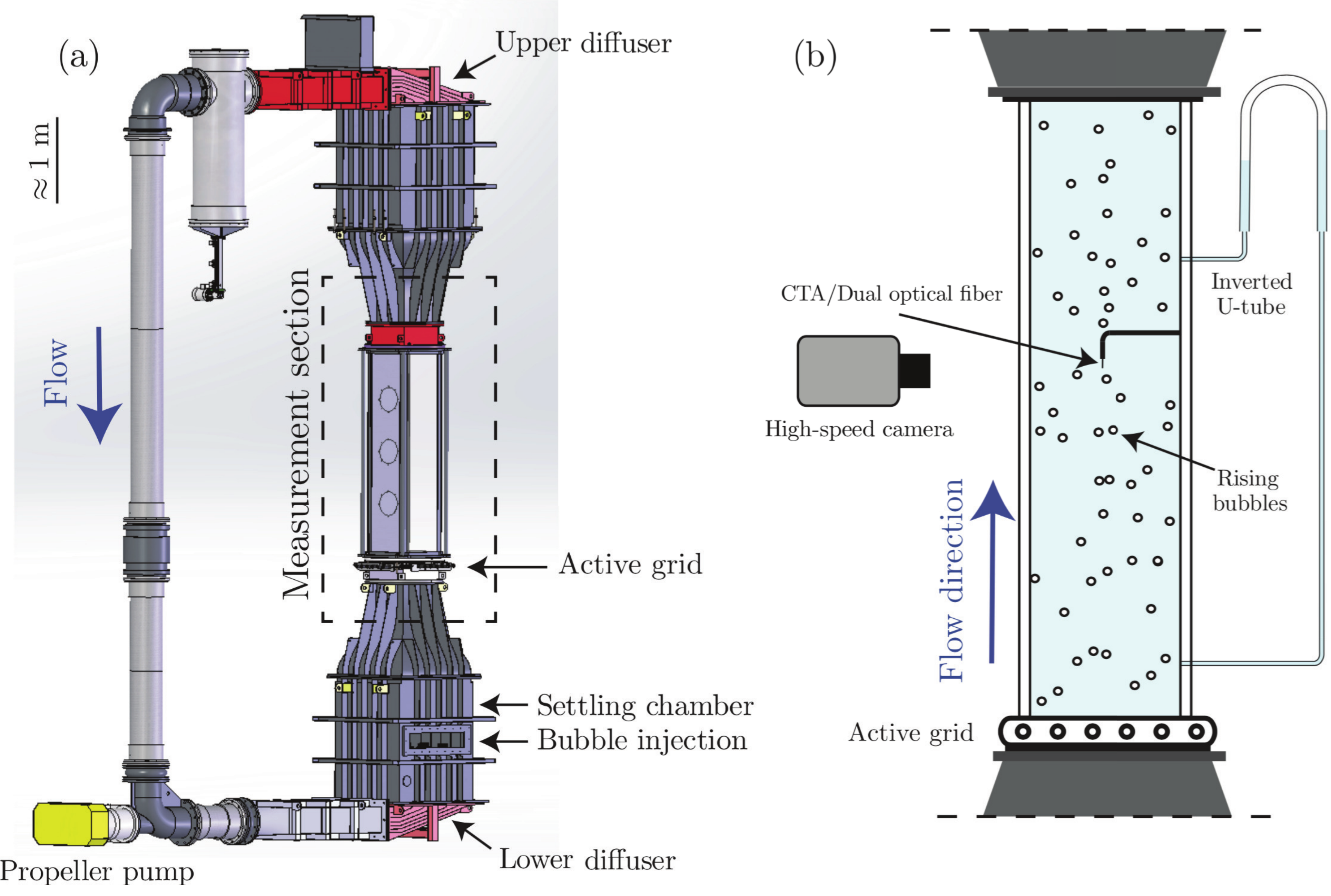}
	\caption{(a)~The Twente Water Tunnel~(TWT) facility, where the experiments were performed. (b)~A zoomed-in view shows the measurement section with the experimental arrangement used in the present study.}
	\label{exp_setup}
\end{figure}

\subsection{Experimental setup}
The experiments are performed in the Twente Water Tunnel, which is a vertical water tunnel as shown in figure~\ref{exp_setup}~(a)~(\cite{poorte2002experiments,mercado2012lagrangian,mathai2016translational}). The measurement section is 2~m high, 450$\times$450 mm in cross-section, and made of three glass walls allowing optical access. An upward mean flow passes through an active grid which is positioned below the measurement section~(figure~\ref{exp_setup}~(b)). The active grid generates nearly homogeneous and isotropic turbulence in the measurement section. By varying the rotation speed of the active grid and the liquid mean flow, different turbulence intensities are achieved. Bubbles are injected below the active grid by means of 69 capillary tubes of inner diameter 0.12~mm. The capillaries are mounted on nine islands which are regularly placed in the settling chamber, below the measurement section, in such a way that the bubble distribution is almost homogeneous in the cross-section of measurement. {\color{black} Indeed, the injection location is about 3m below the measurement location. This gives the bubbles enough time to distributed nearly homogeneously in the cross-section.} The gas volume fraction is varied from 0.25\% to 0.93\% in this work by changing the gas flow rate. An ascending turbulent bubbly flow, {\color{black} rising at a Reynolds number $Re=\frac{V_r d}{\nu}$ (where $V_r$ is the relative rising velocity of the bubble, $d$ the mean bubble diameter, and $\nu$ the cinematic viscosity) ranging from 600 to 900} is thus produced.

\subsection{Instrumentation for the gas-phase characterisation}
The gas phase characterisation consists in measuring the global gas volume fraction, the bubble diameter, and the bubble rising velocity. For that purpose, different measurement techniques have been used. The global gas volume fraction $\alpha$ was measured by using an inverted U-tube~(\cite{rensen2005effect}). {\color{black} We also checked that the global gas  volume fraction was comparable to the one measured locally by an optical probe, indicating that the bubble swarm is distributed homogeneously.} The diameter and the rising velocity of the bubbles were measured by means of a home-made dual optical fiber probe. The optical fiber probe detects the passage of the bubble interface at two measurement points, which are separated vertically by a distance, $\delta =$ 3.41 mm. The bubble velocity is then calculated as $V_r = \delta/\triangle t $, where $\triangle t$ is the time interval between which one bubble collides succesively each probe. Concerning the diameter, it is estimated from the time during which the leading probe is in the gas phase. More details about the signal processing used can be found in \cite{colombet2015dynamics}. 

In order to get more details about the gas phase behaviour, in particular about the aspect ratio of the bubbles, imaging measurements have also been performed by using a high-speed camera (Photron FASTCAM SAX2), with a 105 mm macro lens focused at the center of the measurement section. The recordings are made at an acquisition frequency of 750 fps and with an exposure time of 1/14035~s.  As the diaphragm of the camera is fully open, the depth of field of the camera is relatively short and has been estimated by using a tilted calibration plate. It is around 1~cm. An image-processing method has been developed in order to measure the diameter and the velocity of the bubbles. The bubbles are detected by thresholding based on the spatial intensity gradient of the raw images. In order to detect only single bubbles, two criteria have been imposed on the detected elements, one on the equivalent diameter $d_{eq}=\sqrt{4A/\pi}$ where $A$ is the area of the detected element, and the other on the solidity. \textcolor{black}{The solidity criterion is given by $A/A_{cv}$, where $A$ is the area of the region suspected to be the bubble, and $A_c$ is the area of the smallest convex polygon engulfing the region of the suspected bubble. Here we used a solidity threshold 0.95, which is comparable to the value used in literature~(\cite{honkanen2009reconstruction}).}

\subsection{Instrumentation for the liquid phase characterisation}
The liquid phase has been characterised {\color{black} in the vertical direction only} by means of Constant Temperature Anemometry (CTA) for both single-phase and two-phase flow measurements. The CTA probe is located in the middle of the cross-section of measurement, and the signal is acquired at a frequency of 10 KHz for at least 25 min. It is well known that using CTA in a bubbly flow induces disturbances on the velocity measurement when bubbles collide the CTA probe, generating spurious peaks on the velocity signal. Different methods have been developed to get rid of these spurious peaks and to measure only liquid phase fluctuations~(\cite{van2006energy,martinez2010bubble, rensen2005effect, prakash2016energy}). In the present study, bubble collisions are detected by thresholding the temporal derivative of the raw velocity signal~(\cite{ellingsen1997improvements}). In fact, the temporal derivative of the velocity signal during a bubble collision is typically well over 150 m/s$^2$. For all cases, a velocity derivative of 150 m/s$^2$ was used. This value is well above any liquid phase velocities encountered in the water tunnel flow and therefore, must occur due to the bubble collisions. Thanks to this operation, the liquid phase is distinguished from the bubble collisions. The mean liquid velocity $U$, the standard deviation of the velocity fluctuations $u_{rms}$, and the probability density function (pdf) are thus calculated by considering velocity measurements in the liquid phase only. Before calculating the spectrum of the velocity fluctuations, the gas-phase areas of the signal are removed and replaced by linear interpolation. More details about this operation and its consequences may be found in~\cite{martinez2010bubble,almeras2016time}.

\section{Operating conditions}
\label{sec:op_cond}
\subsection{Turbulence characterisation}
\label{subsec:THI}

\begin{table}
	\centering
	\begin{tabular}{p{0.1\linewidth}p{0.1\linewidth}p{0.1\linewidth}p{0.10\linewidth}p{0.1\linewidth}p{0.1\linewidth}}
		$U$ & $u'_0$ & $Re_\lambda$  &   $\tau_\eta$ & $T_L$ \\
		\hline
		m$/$s & cm$/$s & - &  s &  s \\
		\hline
		0.28  & 2.3 & 177 & 0.083 & 3.6 \\
		0.27  & 3.1 & 242 &  0.066 & 3.9 \\
		0.27  & 3.3 & 262 &  0.061 & 3.9\\
		0.27  & 3.4 & 265 &  0.059 & 3.8\\
		0.47  & 3.5 & 216 &  0.044 & 2.3 \\
		0.46  & 4.6 & 315 &  0.037 & 2.9 \\
		0.46  & 5.1 & 342 &  0.034 & 2.8 \\
		0.46  & 5.5 & 361 &  0.030 & 2.6 \\
	\end{tabular} 
	\caption{Summary of the flow parameters for single-phase measurements. Here, $U$ is the mean flow velocity, $u_0'$ is the standard deviation of velocity fluctuations, $Re_{\lambda}$ is the Taylor Reynolds number, and $\tau_\eta$ and $T_L$ are the dissipative and integral length scales of the turbulent flow, respectively. }
	\label{tab:singlephase}
\end{table}

\begin{figure} \centering
	\includegraphics[scale=0.56]{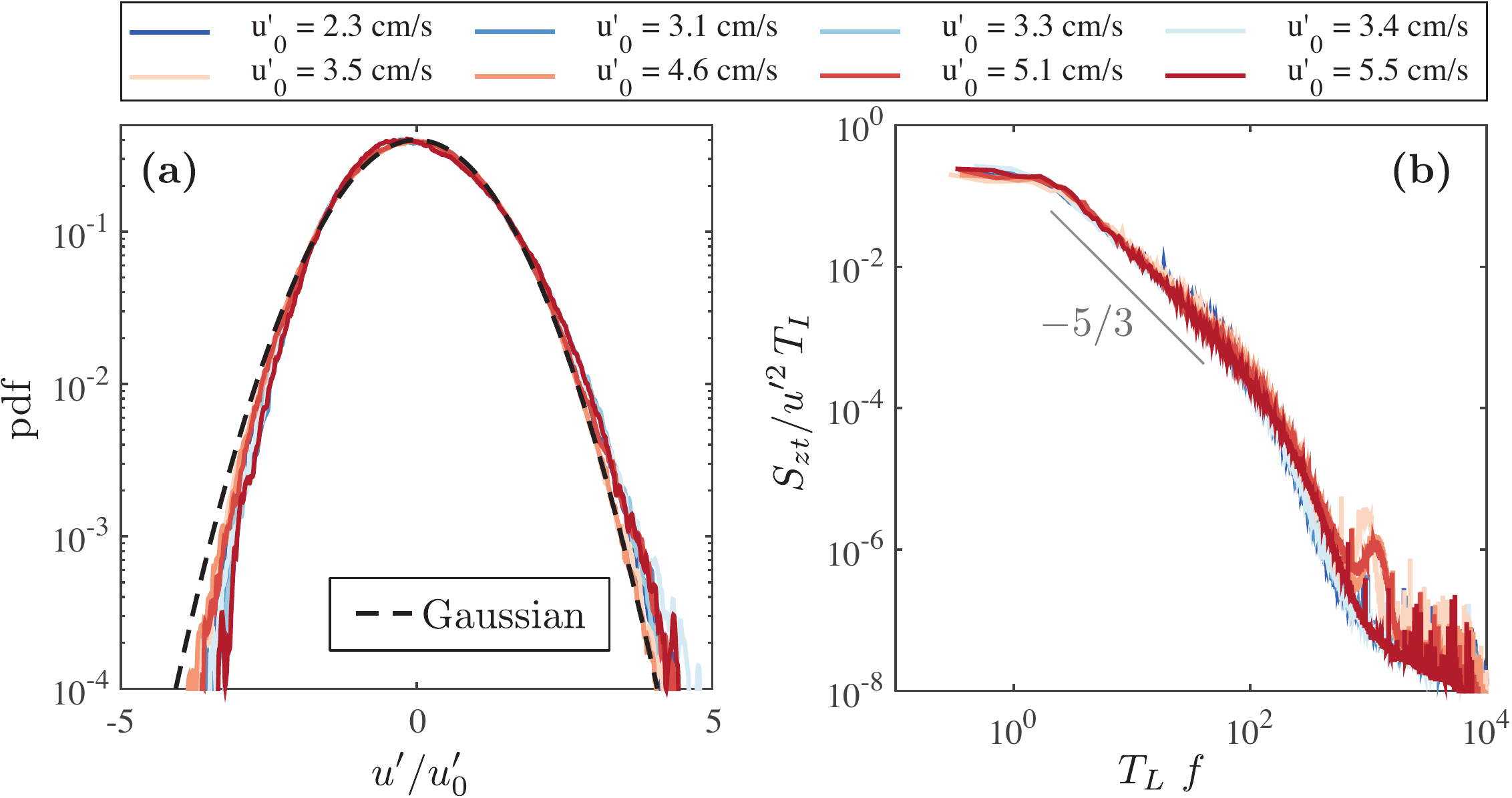}
	\caption{Characterisation of the homogeneous isotropic turbulence for different operating conditions~(single phase flow). (a)~PDF of the velocity fluctuations normalised by the standard deviation. (b)~Normalised spectrum of the velocity fluctuations.}
	\label{fig:pdf_spectrum_singlephase}
\end{figure}

Before investigating turbulent bubbly flows, we will describe the properties of the incident turbulence generated by the active grid. In this study, eight turbulence intensities are investigated, corresponding to two mean flows ($U=0.27$ m/s and $U=0.46$ m/s) and four rotation speeds of the active grid (see table~\ref{tab:singlephase}). Liquid velocity fluctuations are thus increased from $u_0'=2.3$ cm/s to $u_0'=5.5$ cm/s by increasing the mean flow and/or the rotation speed, allowing to vary the Taylor-Reynolds number $Re_\lambda$ from 177 to 361. For the eight cases, the turbulent velocity fluctuations are nearly homogeneous and isotropic, as also seen in pevious studies using the same setup~(\cite{prakash2016energy,mathai2015wake,martinez2010bubble}). In fact, once normalised by the corresponding standard deviation, the pdf of the velocity fluctuations shows a Gaussian behaviour, irrespective of the level of turbulence~(figure~\ref{fig:pdf_spectrum_singlephase} (a)). The spectrum of the velocity fluctuations presents a $-5/3$ scaling over two decades (figure~\ref{fig:pdf_spectrum_singlephase}~(b)), within the frequency range [$1/T_L ; 1/\tau_\eta$] (table \ref{tab:singlephase}), where $T_L$ is the integral time scale calculated from the dissipation rate $\epsilon$ ($T_L=(\frac{3}{4} C_0 \frac{\epsilon}{k})^{-1}$, with $k=3/2 u_0'^2$, $C_0 = 2.1$). The energy dissipation rate $\epsilon$ is estimated from the second order longitudinal structure function, which is calculated by means of the Taylor hypothesis, similar to the methods followed in \cite{mathai2015wake,mathai2016microbubbles}. The dissipative time scale $\tau_\eta \equiv (\nu/\epsilon)^{1/2}$ is finally estimated for each turbulence level~(table \ref{tab:singlephase}). {\color{black} It must be noted that the above procedures involve Taylor's frozen flow hypothesis and the assumption that the second order structure function~$D_{LL}$ = $C_2  (\epsilon r)^{2/3}$~(\cite{pop00}). These assumptions are reasonable for the range of parameters in the present study. Typical errors in these estimates lie within 20\%, and have been quantified in prior investigations~(\cite{poorte2002experiments,mercado2012lagrangian}).}

\subsection{Bubble properties}

\begin{figure} \centering
	\includegraphics[scale=0.52]{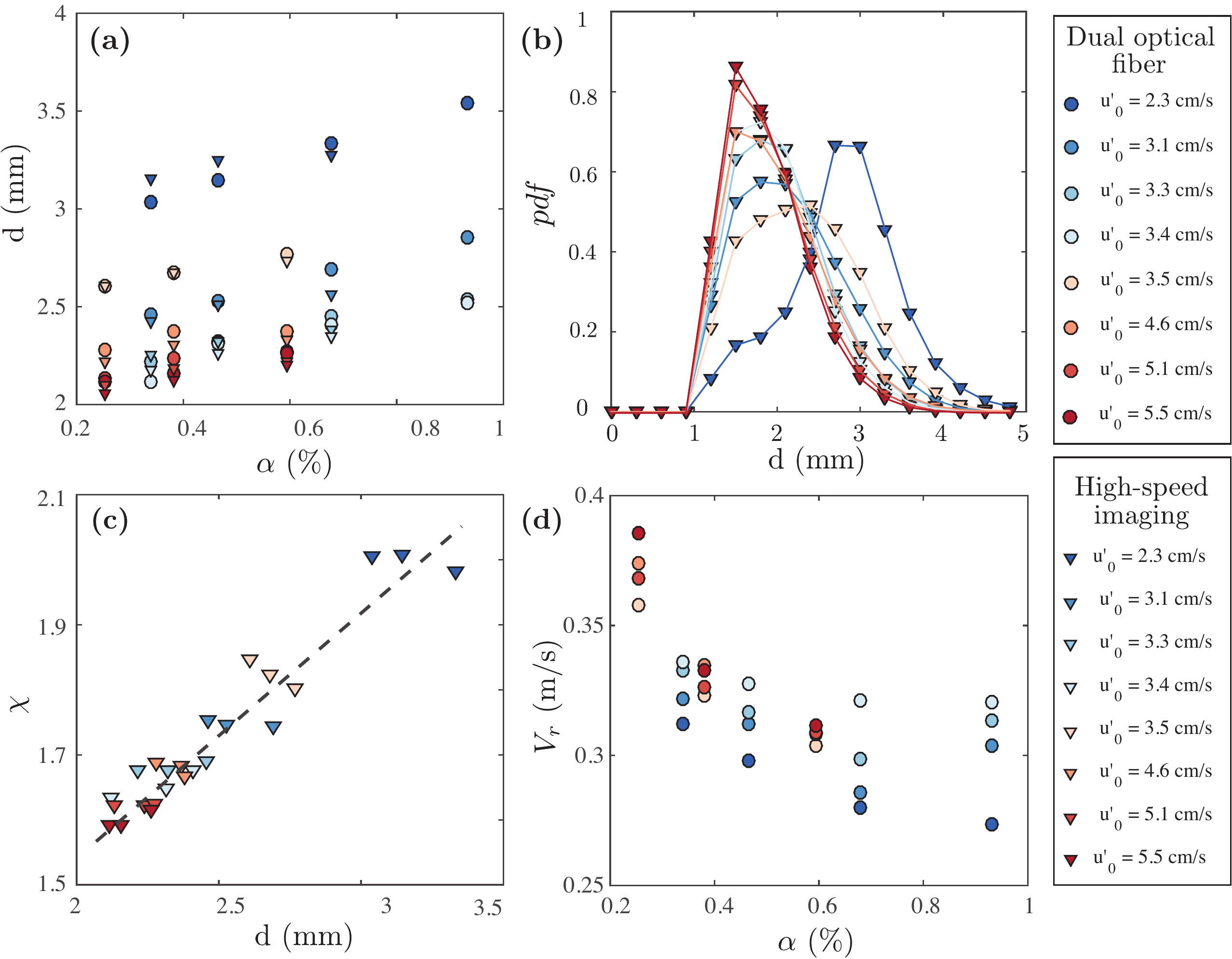}
	\caption{Gas phase charaterisation for different turbulence intensities. (a)~Mean bubble diameter $d$ as a function of the gas volume fraction $\alpha$ for different turbulence intensities $u'_0$. (b) Bubble size distribution for different turbulence levels at the lowest gas volume fraction in our experiments. (c) Aspect ratio $\chi$ of bubbles vs mean bubble diameter $d$ , showing a nearly linear dependence. (d)~Relative bubble velocity as a function of the gas volume fraction $\alpha$ for different turbulence intensities. The circle symbols denote the measurements made from the dual optical probe, and the triangle symbols denote estimates obtained from high-speed imaging. The color code reflects the rms of liquid velocity fluctuations $u_0'$ of the incident turbulence.}
	\label{fig:bubble_diam}
\end{figure}

For the eight turbulence levels considered in this study, the gas volume fraction is varied from 0\% to 0.93 $\%$ for $U=0.27$ m/s and from 0\% to 0.59 $\%$ for $U=0.46$ m/s. The mean bubble diameter ranges from 2.0 mm to 3.6 mm and depends both on the gas volume fraction and the turbulence level~(see figure~\ref{fig:bubble_diam} (a)). For all levels of turbulence, the mean bubble diameter increases with the gas volume fraction. Moreover, a strong influence of the rotation speed of the active grid on the bubble diameter can be observed : the faster the rotation speed, the smaller are the bubbles. This trend has already been observed by \cite{prakash2016energy}. \textcolor{black}{The size distributions of the bubbles diameter are presented in figure~\ref{fig:bubble_diam} (b) for every level of turbulence at the lowest gas volume fraction investigated. The standard deviation of the bubble diameter is about $25 \% - 30$\% of the mean diameter for all the gas volume fractions and turbulence levels studied. Concerning the aspect ratio of the bubbles, it ranges from 1.6 to 2.0, depending mainly on the rotation speed of the grid and weakly on the gas volume fraction.} As shown in figure \ref{fig:bubble_diam} (c) the aspect ratio of the bubble is fully controlled by the bubble diameter: smaller bubbles are more spherical compared to larger ones. The Weber number, $\text{We} = \frac{\rho V_r^2 \ d}{\sigma}$ in the shown range varies from 1.8~(nearly spherical) to 7.1~(considerably deformed).

The relative bubble rise velocity $V_r$ {\color{black} measured by optical fiber} is shown in figure~\ref{fig:bubble_diam}(d) as a function of the gas volume fraction for the eight turbulence levels studied. {\color{black}We checked that the relative velocity measured by the dual optical probe was comparable to the one evaluated from the high-speed images (not shown here). The variation between both techniques was within $\pm 5 \%$ for the moderate volume fractions studied here.} The rising velocity is calculated as the difference between the measured bubble velocity $V_b$ and the liquid mean flow $U$. Under the present operating conditions, the relative rise velocity ranges from 0.27 m/s to 0.39 m$/$s and depends both on the turbulence level as well as the gas volume fraction. For any chosen turbulence intensity, the relative rise velocity decreases with the gas volume fraction, as has already been observed for pseudo-turbulence~(\cite{riboux2010experimental,colombet2015dynamics}). However, more experiments would be required to have a deeper understanding on the dependance of $V_r$ on the gas volume fraction and the turbulence level, which is outside the scope of this study. Nevertheless, we measure precisely the relative rise velocity for the present operating conditions, since this is a key parameter affecting the liquid motion.

\section{Dynamics of the liquid phase within the swarm}
\label{sec:dyn_liq_Ncond}
 
\begin{figure} \centering
	\centering
	\includegraphics[scale=0.5]{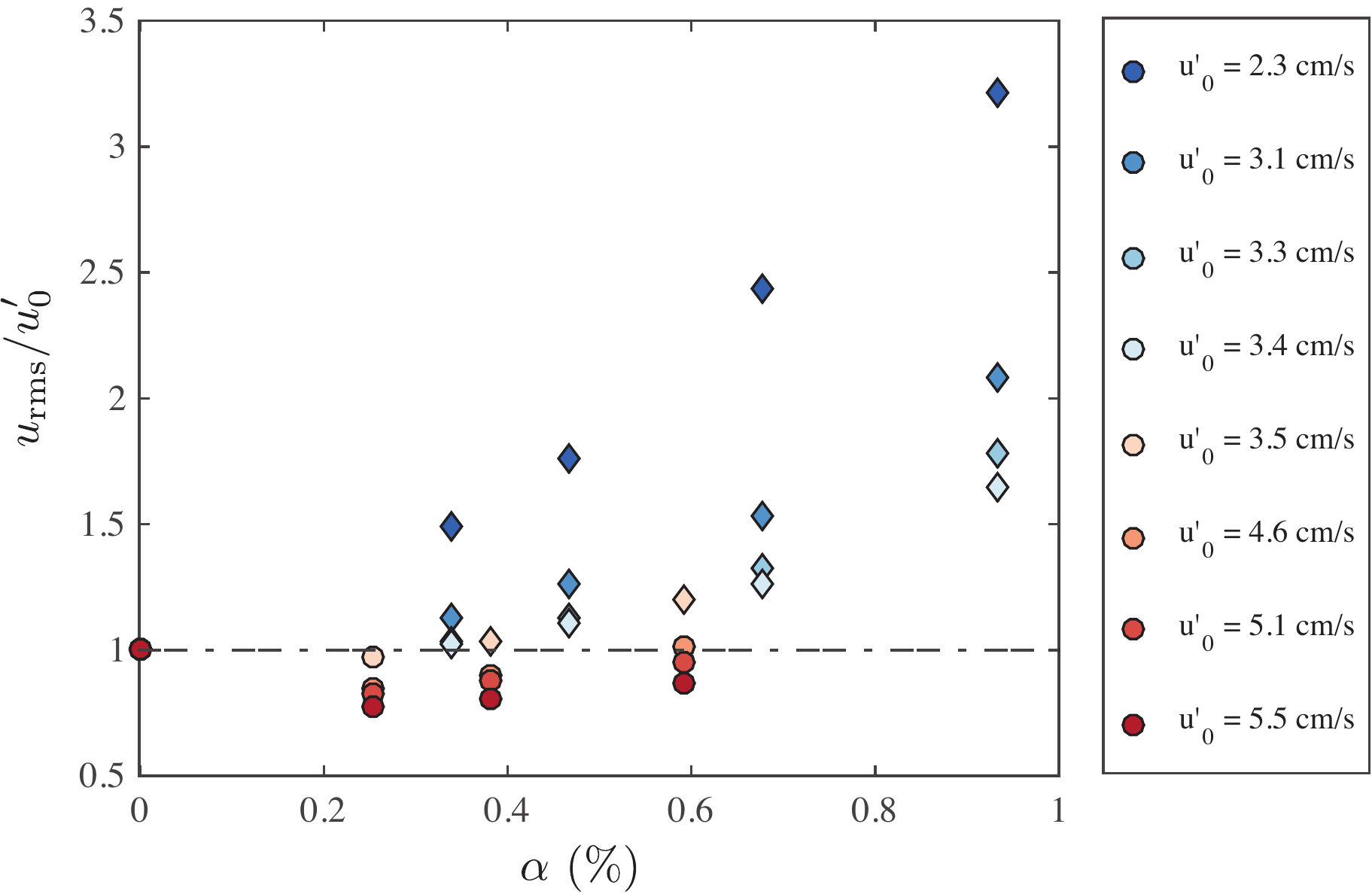}
	\caption{Standard deviation of the liquid velocity fluctuations in the vertical direction normalised by $u'_0$ as a function of the gas volume fraction for different levels of turbulence.}
	\label{fig:var_vs_alpha}
\end{figure}


\begin{figure} \centering
	\includegraphics[scale=0.5]{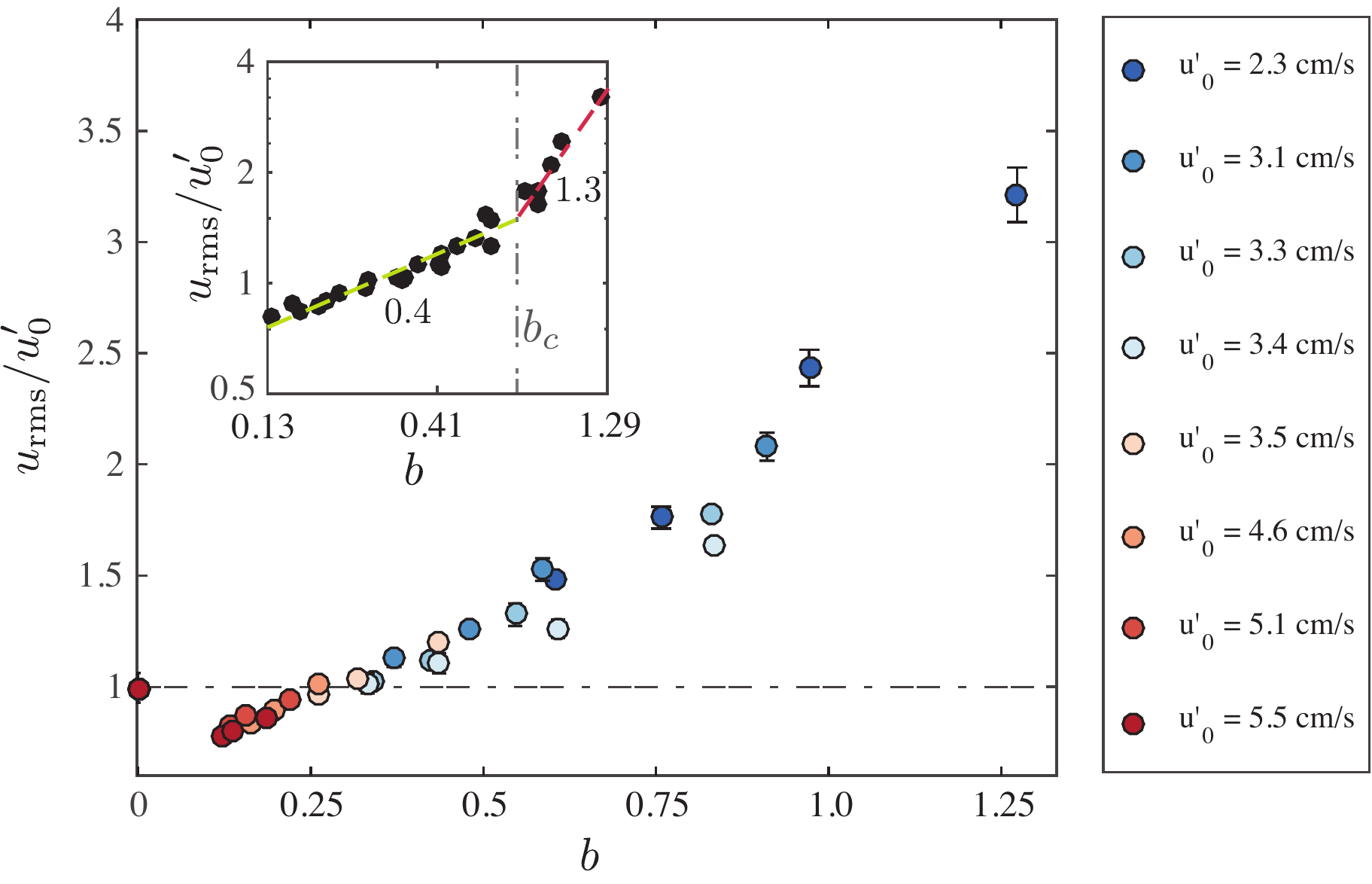}
	\caption{{\color{black} Standard deviation of the liquid velocity fluctuations in the vertical direction normalised by $u'_0$ as a function of the bubblance parameter. The errorbars denote the minimum and maximum values of five equally sampled datasets. Inset shows the same plot on log-log scale.}}
	\label{fig:var_vs_bubblance}
\end{figure}

\subsection{Variance of the velocity fluctations} 
Figure \ref{fig:var_vs_alpha} shows the standard deviation of the velocity fluctuations $u_{rms}$ normalised by the incident turbulent fluctuation $u'_0$ as a function of the gas volume fraction $\alpha$ for different levels of turbulence. We notice that even after normalisation by the fluctuation of the incident turbulence, the standard deviation of the velocity fluctuation still shows a strong dependence both on $\alpha$ and on $u'_0$. In particular, for $u'_0<3.5$ cm/s, the velocity fluctuations are larger than the one produced by the incident turbulence and increase with $\alpha$ whereas they are lower for $u'_0>4.6$ cm/s.{\color{black} These two behaviours can be explained by introducing the bubblance parameter $b$, which compares the energy of the fluctuations produced by the bubble swarm to the one produced by the incident turbulence ${u'_0}^2$. However, predicting theoritically the absolute level of energy produced by a bubble swarm is a difficult task. Following a theortical approach based on potential theory, it has been shown that this energy is equal to $C_m V_r^2 \alpha$, where $C_m$ is the added mass coefficient (\cite{lance1991turbulence,van1998pseudo}). Experimental work from \cite{riboux2010experimental} indicates that the variance of the velocity fluctuations produced by a bubble swarm is equal to $\gamma^2 V_r^2 \alpha$, where $\gamma$ is a prefactor which takes into account the anisotropy of the velocity fluctuations in the vertical and horizontal directions ($\gamma = 1.94$ for the vertical direction). Even if some discussions about the absolute level of the energy produced by a bubble swarm are still going on, it is reasonable to consider that the energy produced by a bubble swarm is proportional to $V_r^2 \alpha $. We thus decided to express the bubblance parameter as:
\begin{equation}
b = \frac{V_r^2 \alpha}{{u'_0}^2}.
\label{eq:def_bubblance}
\end{equation}
Consequently, $b=0$ corresponds to single phase flow, whereas $b =\infty $ is for pseudo-turbulence cases. Note that the above definition of the bubblance parameter differs from prior definitions in \cite{rensen2005effect,prakash2016energy} by the factor $C_m=\frac{1}{2}$. The added mass coefficient $C_m = \frac{1}{2}$ comes from potential flow theory for spherical bubbles, while in the present case, we have a swarm of high Reynolds number deformable bubbles. Therefore, we chose to define the bubblance parameter without this prefactor~(Equation~\ref{eq:def_bubblance}), since this is enough to capture the ratio of the energy of the bubble swarm to that of the incident turbulence. Regardless of the definition chosen, the prefactor does not influence the main conclusions of the present study.}\\

The strong correlation between the bubblance parameter and the intensity of the velocity fluctuation is particularly evident from figure \ref{fig:var_vs_bubblance}, in which the velocity fluctuations $u_{rms}$ normalised by $u'_0$ are plotted as a function of $b$. Regardless of the level of turbulence, all data points collapse on a master curve, which shows a non-monotonic behaviour with the bubblance parameter. For $b<0.27$, the velocity fluctuations are smaller than the one produced by the incident turbulence. Thus, adding a small amount of bubbles in a very strong incident turbulence deeply modifies the nature of the flow, leading to a reduction in the intensity of the velocity fluctuations, or turbulence attenuation~(\cite{cisse2015turbulence,mazzitelli2003effect}). However, for $b>0.27$, an enhancement of the velocity fluctuations compared to single phase flow is observed. Note that the lack of data in the range $0<b<0.13$ limits us from having a complete description of the transition between a single-phase flow and a turbulent bubbly flow evolving at low bubblance parameter. It is however difficult to investigate this range of $b$, since it requires smaller gas volume fractions. The current capillary islands in our water tunnel could not produce a homogeneous bubble swarm at such low gas flow rates.\\

We will thus now focus only on the turbulent bubbly flows corresponding to $b>0$. In this regime, the normalised velocity fluctuations increase monotonically with the bubblance parameter. Two regimes can be observed, separated by a critical bubblance parameter $b_c \approx 0.7$~(see the inset to figure~\ref{fig:var_vs_bubblance}). For $b<b_c$, the normalised velocity fluctuations evolve roughly as  $u_{rms}/u_0' \propto b^{0.4}$, whereas they increase much faster for $b>b_c$, approximately as $\propto b^{1.3}$. These two regimes can be related to a stronger contribution of the bubble wakes when the bubblance parameter increases, which will be discussed in section \ref{sec:cond_properties}.

\subsection{PDF of the velocity fluctuations}
\begin{figure} \centering
	\centerline{\includegraphics[scale=0.55]{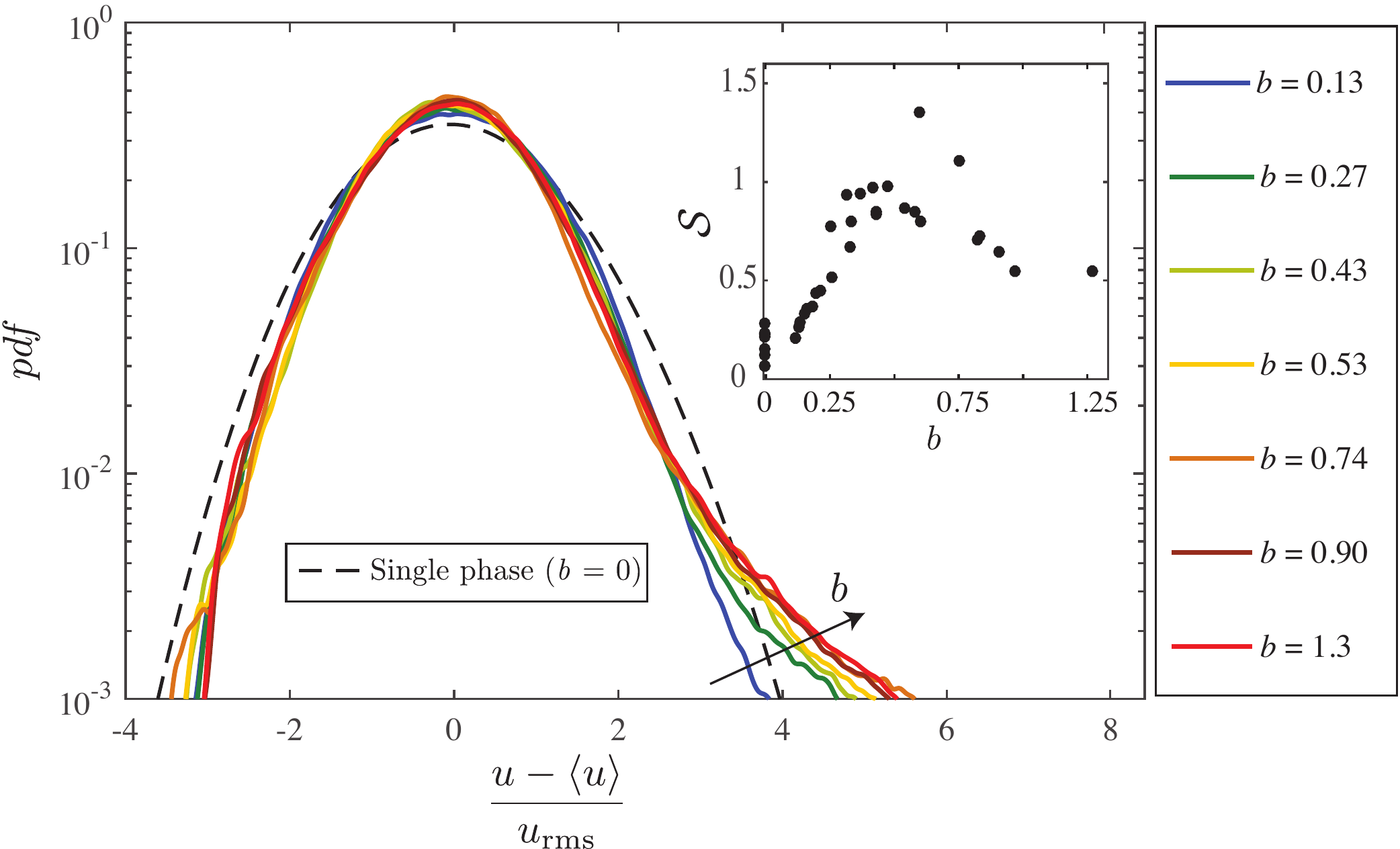}}
	\caption{PDF of the liquid velocity fluctuations normalised by $u_{rms}$ for different bubblance parameters. The skewness~$\mathcal{S}$ of the pdf for increasing $b$ is shown as inset.}
	\label{fig:pdf_liq}
\end{figure}

We will now discuss how the structure of the velocity fluctuations depend on the bubblance parameter $b$. For that purpose, the pdf of the vertical velocity fluctuations normalised by the standard deviation $u_{rms}$ is plotted in figure~\ref{fig:pdf_liq} for $b$ in the range $0 - 1.3$. For $b=0$, which corresponds to single phase flow, the pdf of the normalised velocity fluctuations shows nearly Gaussian behaviour. For turbulent bubbly flows ($b>0$), a skewness appears with stronger fluctuations for positive values but the pdf remains nearly Gaussian for negative and more probable values. Interestingly, the pdfs do not exhibit exponential tails, as would be expected for pseudo-turbulence ($b=\infty$). It has been shown by \cite{riboux2010experimental} that the exponential tails are mainly due to the agitation produced by the wakes in interaction. In a turbulent bubbly flow, the agitation does not result solely from the interactions of the wakes, since the incident turbulence plays a role as well. \\

The positive skewness of the pdf is also observed experimentally for different operating conditions and is known to be the signature of the bubble wakes (\cite{prakash2016energy, riboux2010experimental}). However, in \cite{prakash2016energy}, they could not observe a clear trend with respect to the $b$ parameter. Here we demonstrate that by increasing the bubblance parameter, the positive skewness increases to a maximum value for $b \approx 0.7$~(inset to figure~\ref{fig:pdf_liq}). This can be interpreted as a stronger contribution of the bubble wakes as the bubblance parameter increases. Beyond $b \approx 0.7$, $\mathcal{S}$ appears to decrease a bit before saturating at large values of the bubblance parameter.

\subsection{Frequency spectrum of the velocity fluctuations}

\begin{figure} \centering
	\centerline{\includegraphics[scale=0.65]{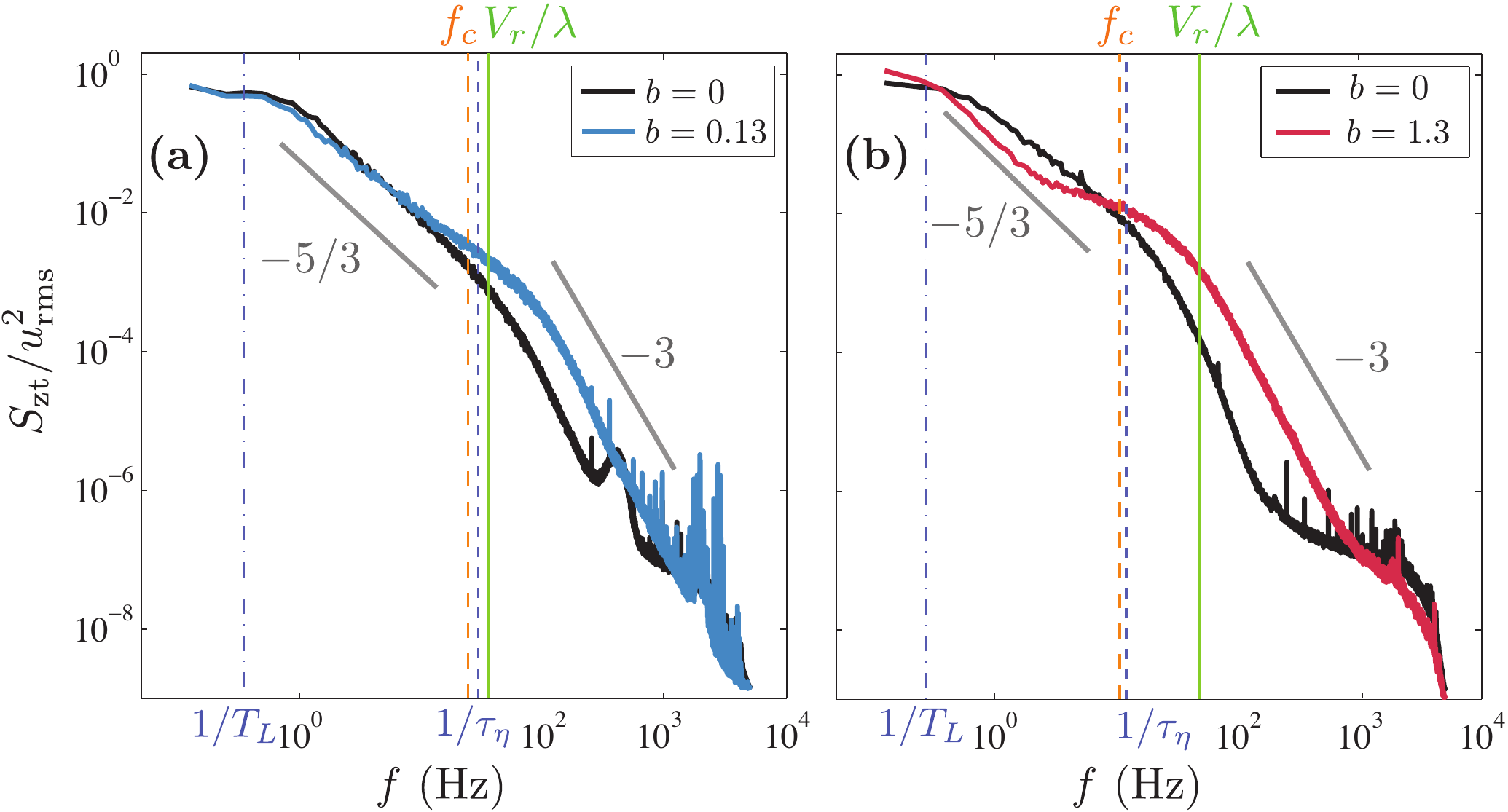}}
	\caption{Energy spectra of the velocity fluctuations for (a)~$b = 0.13$ and (b)~$b = 1.3$. The black curves in (a) and (b) denote the spectrum for the single-phase case before the addition of bubbles.}
	\label{fig:spe_SP_TP}
\end{figure}

\begin{figure} \centering
	\centerline{\includegraphics[scale=0.55]{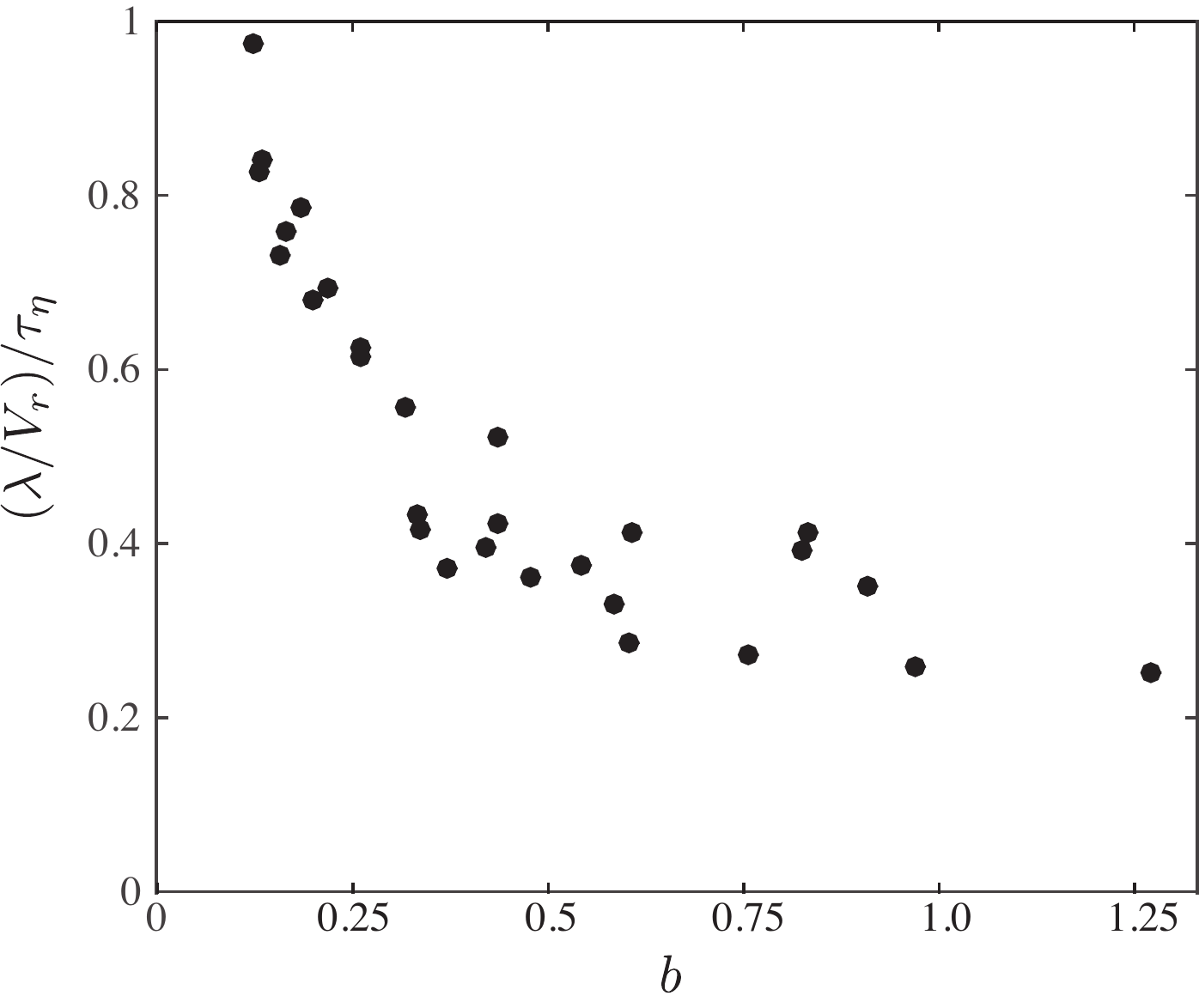}}
	\caption{Ratio of the integral time scale~($\lambda/V_r$) imposed by the bubbles over the dissipative time scale ($\tau_\eta$) in the flow as a function of the bubblance parameter $b$.}
	\label{fig:lambda_over_tnu}
\end{figure}

\begin{figure} \centering
	\centerline{\includegraphics[scale=0.55]{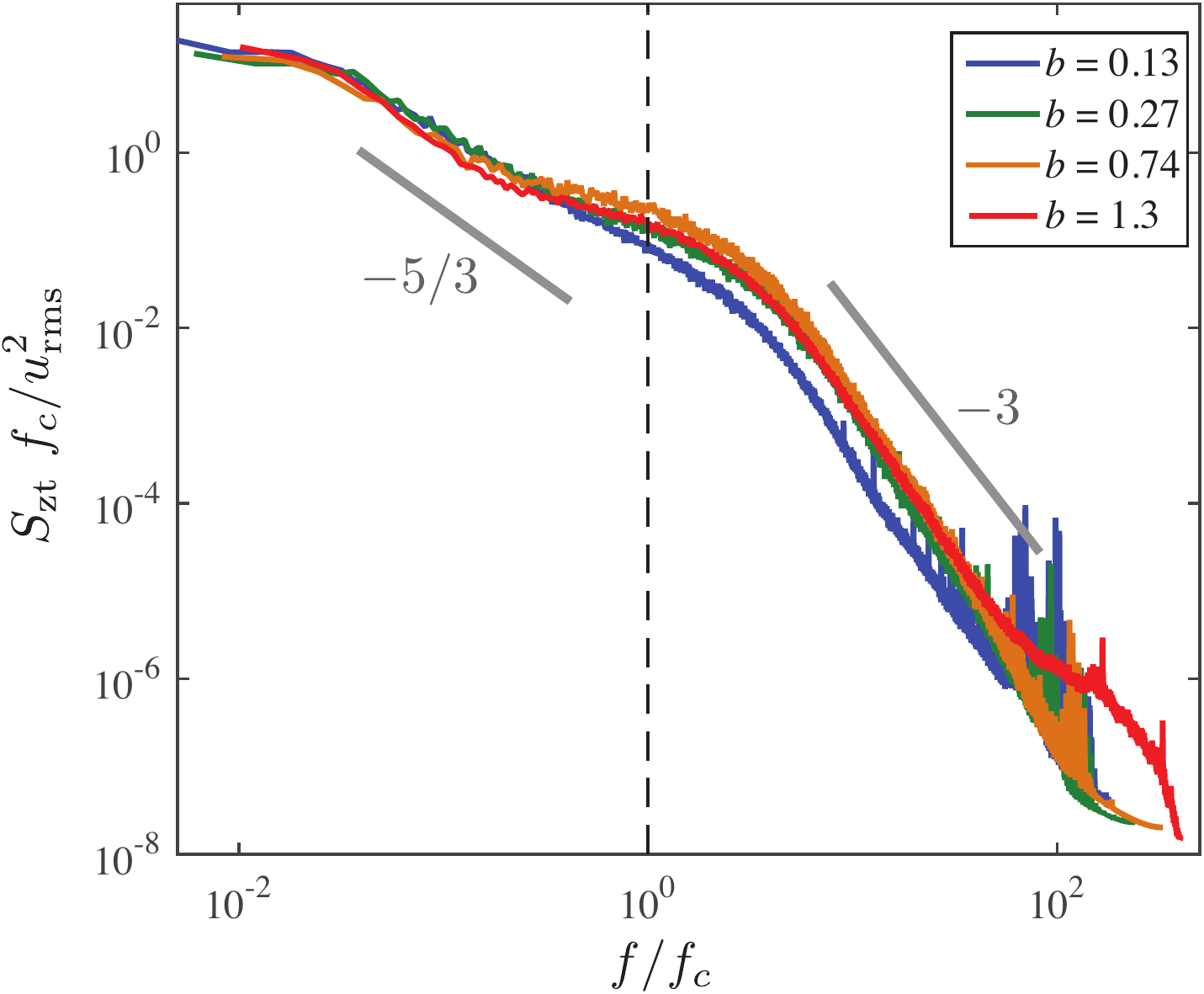}}
	\caption{Energy spectrum of the velocity fluctuations normalised by $f_{c}$ for different values of the bubblance parameter $b$.}
	\label{fig:spe_norm_fcwi}
\end{figure}

We now investigate the frequency spectrum $S_{zt}$ of the velocity fluctuations for different values of $b$. For each operating condition, the spectrum is calculated by averaging at least 180 spectra, each spanning a recording time duration of 8~s, and normalised by $u_{rms}^2$. First, the two-phase flow spectrum is compared to the corresponding single phase one, for $b=0.13$ (figure~\ref{fig:spe_SP_TP}(a)) and $b=1.3$ (figure \ref{fig:spe_SP_TP}(b)). As already seen in subsection \ref{subsec:THI}, the classical $-5/3$ decrease can be observed for single phase flow in the range of frequency $[1/T_L; 1/\tau_\eta]$.  {\color{black}Beyond this range, for $f > 1/\tau_\eta$ lies the viscous dissipative range with a very low energy content. The higher negative slope in the dissipative range is expected, and this is likely the range prior to the classical exponential decay~\citep{pop00}.}

For the turbulent bubbly flows, the $-5/3$ scaling is still observed for the frequency range $[1/T_L; 1/\tau_\eta]$, and is then followed by $-3$ scaling for higher frequencies. {\color{black}When bubbles are added to the system, there is a significant increase in the energy of the flow, as was already seen in the $u_{rms}$ vs $b$ plot in figure~\ref{fig:var_vs_bubblance}.  As is known from~\cite{riboux2010experimental,lance1991turbulence}, the $-3$ slope is the signature of the bubble-induced turbulence and develops for length scales smaller than $\lambda=d/C_{d0}$, where $C_{d0}$ is the drag coefficient of a single rising bubble.} The characteristic time scale $T_{pseudo}$ imposed by the bubble swarm may then be calculated as $T_{pseudo}=\lambda/V_r$, and the corresponding frequency scales are shown by the green lines in figure~\ref{fig:spe_SP_TP}(a) \& (b). Irrespective of the bubblance parameter, we observe that the $-3$ scaling subrange starts at the frequency $1/T_{pseudo}=V_r/\lambda$. Since the frequency $V_r/\lambda$ imposed by the bubbles is larger than the Kolmogorov frequency 1/$\tau_\eta$ for the range of bubblance parameters studied here~(figure \ref{fig:lambda_over_tnu}), we can observe a distinct separation of the time scales in the flow, i.e the incident turbulence acting for the time scales larger than $\tau_\eta$, and the bubble-induced turbulence for time scales lower than $\lambda/V_r$, which results in two distinct slopes. However, for $b=1.3$, the spectrum is deformed in the $-5/3$ subrange and presents a bump at $f_{c} \equiv 0.14V_r/d \approx 12$~Hz~(figure~\ref{fig:spe_SP_TP}(b)). {\color{black}Beyond this frequency, we again see the $-3$ slope characteristic of pseudo-turbulence. It may be noted that the steep slope for $f > 1/\tau_\eta$ in the single phase cases is distinctly different from the $-3$ slope of the  two-phase cases. This is clear from the fact that the $-3$ slope of two-phase flow has much higher energy content, and occurs at larger frequencies as compared to the single-phase cases.}

%

Figure~\ref{fig:spe_norm_fcwi} presents the spectrum of the velocity fluctuations normalised by $f_{c}$ for different bubblance parameters. It appears that the bump around $f_{c}$ is present only for $b > 0.27$. This excitation frequency has already been observed by \cite{riboux2013model} for pseudo-turbulence. The authors attributed it to the injection of energy by the collective wake instability into the flow. However, this frequency is also comparable to the vortex shedding frequency $f_v \equiv St \ V_r/d$ of a single bubble at $Re_b \approx 500$, where the Strouhal number $\text{St} \approx$ 0.13~(\cite{wu2002experimental,shew2006force}). It is important to note that this frequency is also close to the one defined by \cite{prakash2016energy} as $f_b=\frac{V_r}{2\pi d}$. We note that the time-scale separation in the present experiments is too narrow to reveal the true origin of this frequency, but this is an interesting issue for future investigation.

\section{Conditional properties of the liquid phase}
\label{sec:cond_properties}
\subsection{Velocity disturbance behind the bubbles}
\begin{figure} \centering \centering
	\centerline{\includegraphics[scale=0.55]{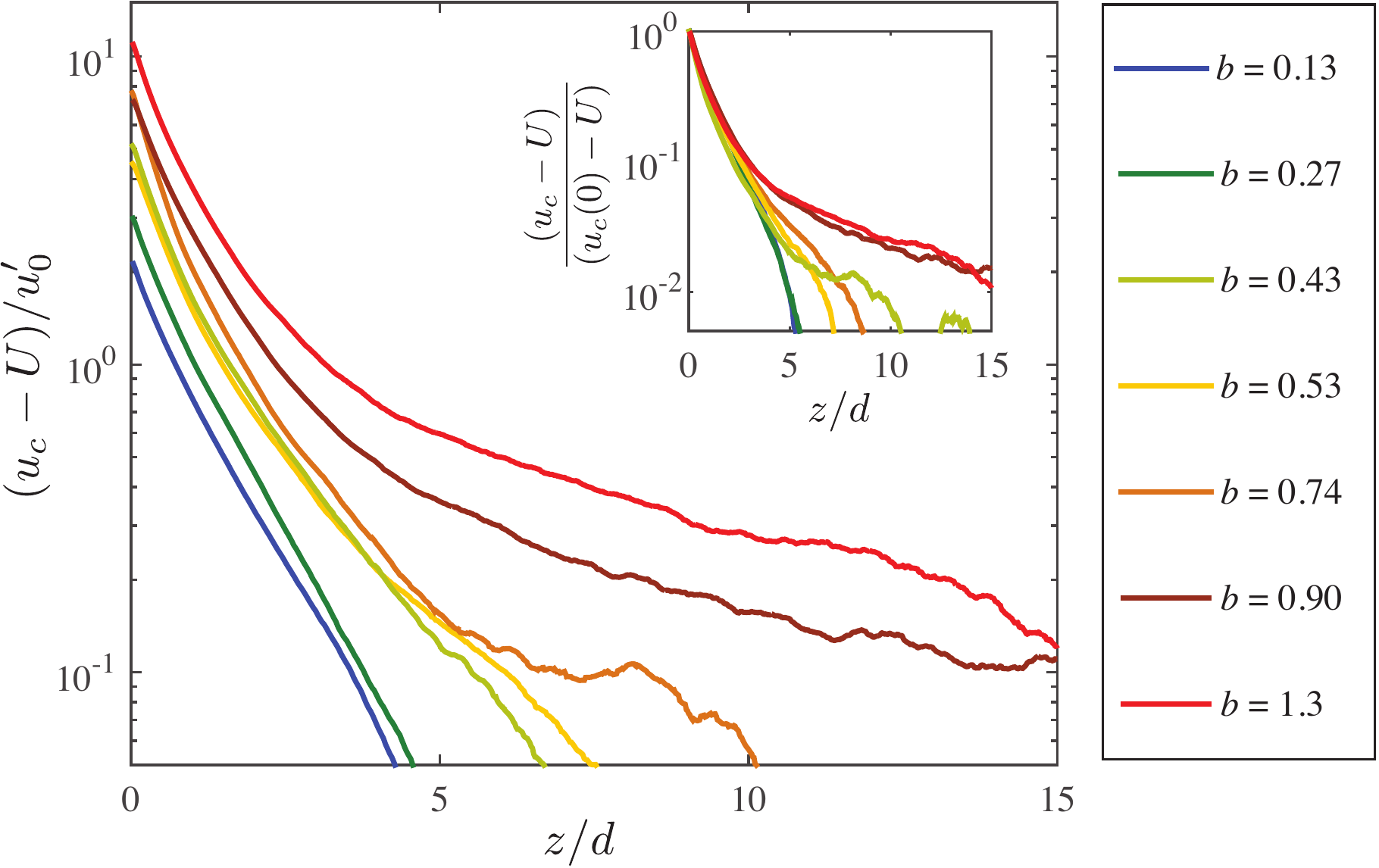}}
	\caption{Velocity disturbance behind a bubble in the turbulent bubbly flow. Inset shows the velocity disturbance normalised by $(u_c(0) - U)$, which reveals that the primary wake region is not affected by $b$.}
	\label{fig:wake}
\end{figure}

We will now study the velocity disturbance behind individual bubbles. The conditional velocity $u_c$ is calculated by averaging the velocity disturbance just behind each detected bubble (at least 1087 bubbles were detected for each measurement point). It is important to mention that the neighbouring bubbles could induce pertubations on the conditional velocity, mainly through their upstream perturbations. The upstream disturbance induced by the next bubble is thus excluded by shortening the detected sample $3\frac{d}{V_b}$ s before the following bubble is detected. \textcolor{black}{The three diameters distance was chosen based on the analysis of \cite{roig2007measurement}, the influence of the upcoming bubble was seen to be minor. We thus checked that in the present study, the upstream perturbation doesn't exceed three bubble diameter too.} Finally, Taylor hypothesis is used to convert the temporal signal to a spatial one by using the mean rising velocity of the bubbles $V_b$ (\cite{hinze1975turbulence}). The velocity disturbance behind the bubbles normalised by the turbulent fluctuations $u'_0$ is shown in figure~\ref{fig:wake} for different bubblance parameters. It is clear that the bubblance parameter strongly affects the structure of the wake. Both the length of the wake and its shape depend on the bubblance parameter. For low bubblance parameter ($b<0.7$), the wake is around 5$d$ long and presents a single exponential decrease. However, when the bubblance parameter is larger than 0.7, the wake length considerably increases, reaching almost 15$d$ to 20$d$. Moreover, the conditional mean velocity presents an exponential decrease with a change in slope at a critical distance $z_c$ at which $u_c(z_c)=u'_0$. We can thus distinguish two domains in the wake: a primary wake for $z<z_c$, referred to as PW, and a secondary wake (SW) for $z>z_c$.


\begin{figure} \centering \centering
	\centerline{\includegraphics[scale=0.55]{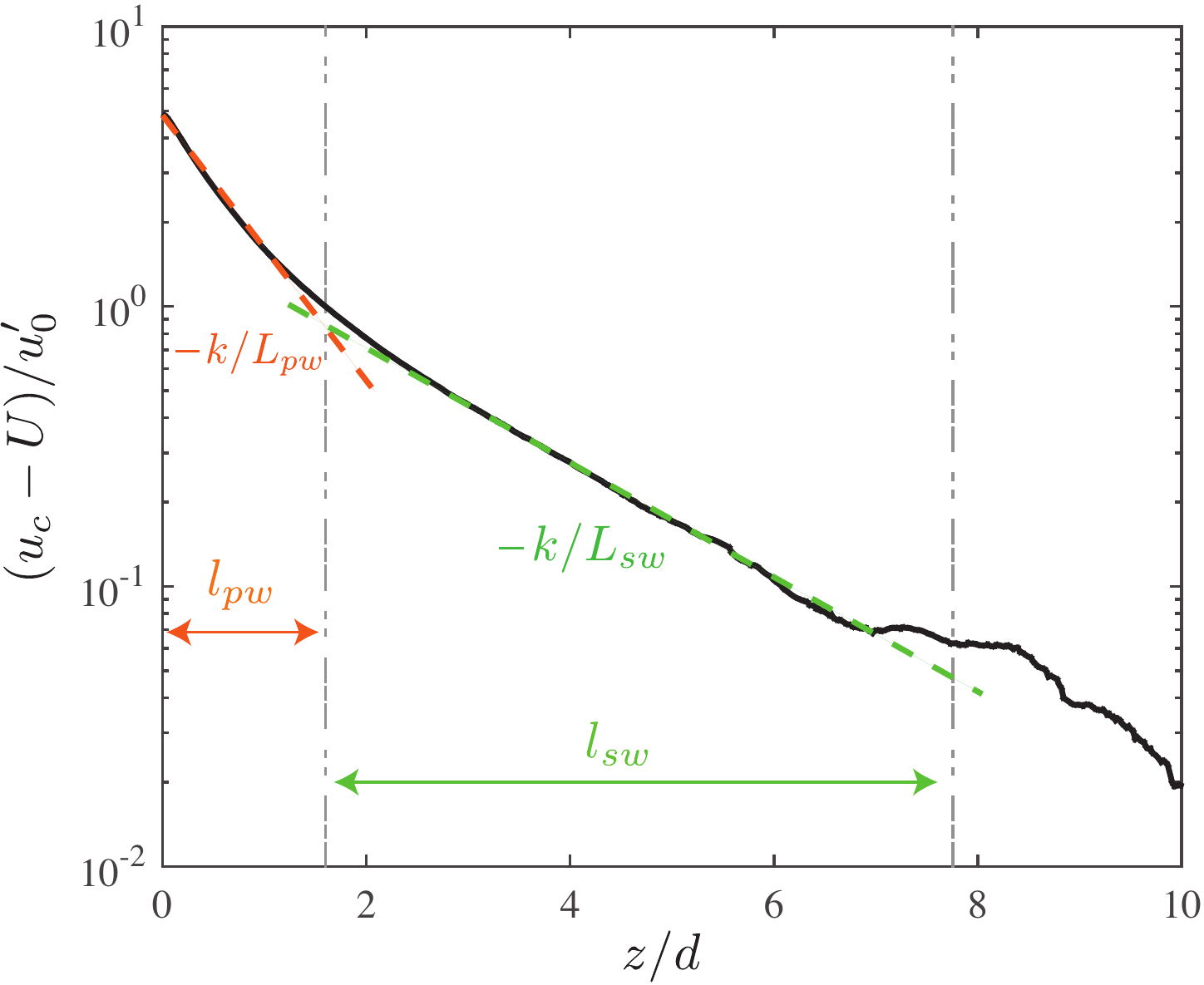}}
	\caption{Classification of the bubble-wake into primary~(PW) and secondary wake regions~(SW). Here $k = 1/ \ln (10)$ is a prefactor in the slope shown in the above plot.}
	\label{fig:wake_structure}
\end{figure}

A primary wake~(PW) and a secondary wake~(SW) were also observed by \cite{legendre2006wake} for a single bubble in a turbulent pipe flow. The PW was defined as the region which showed a fast velocity decrease, and the transition from PW to SW occured when the velocity deficit was of the order of the incident turbulence.
Following a similar approach, we quantify the PW and the SW in our case by defining two characteristic length scales each:
a characteristic length $l$ and a characteristic decay length $L$~(figure \ref{fig:wake_structure}). In the primary wake ($0<z<l_{pw}$), the velocity disturbance is well described by an exponential decrease $u_c(z)-U = (u_c(0)-U) \times exp(-\frac{z}{L_{pw}})$, where $L_{pw}$ is the characterstic decay length of the primary wake. The characteristic length, $l_{pw}$, is calculated by using the exponential fitting as $u_c(z=l_{pw})=u'_0$. {\color{black} Concerning the secondary wake region, which is defined in the region $l_{pw}<z<l_{sw}$, the velocity disturbance follows also an exponential decrease, which can be expressed as $u_c(z)~-~U~=~(u_c(l_{pw})~-~U)~\times~exp(-\frac{z-l_{pw}}{L_{sw}})$, where $L_{sw}$ is the characteristic decay length of the secondary wake, and $l_{sw}$ is the length of the secondary wake. The length of the secondary wake $l_{sw}$ is defined from the exponential fitting for $u_c(z)-U=0.05 \ u_0'$. The threshold was chosen as low as possible so that the length of the secondary wake was not underestimated while not taking into account the eventual noise present in the mean velocity deficit far from the bubble interface. In this regard, the value $0.05 \ u_0'$ seemed to be a reasonable choice. For the fitting of the secondary wake~(see figure~\ref{fig:wake_structure}), we checked manually for all the cases that the fits were reasonable. We expect the uncertainties to be within 20$\%$.}

\begin{figure} \centering \centering
	\centerline{\includegraphics[scale=0.6]{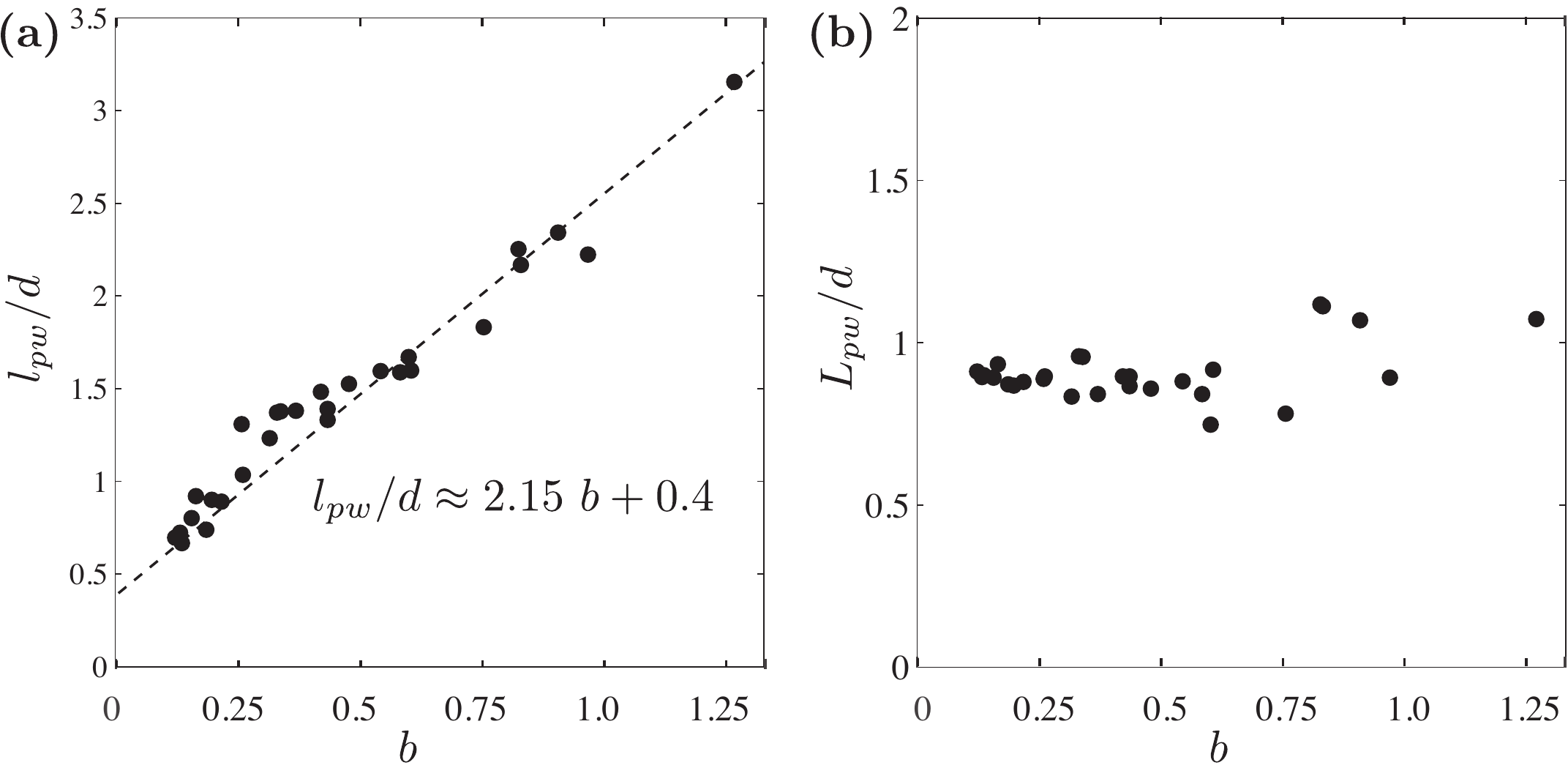}}
	\caption{Characteristics of the primary wake. (a)~Characteristic length. (b)~Characteristic decay length.}
	\label{fig:cara_near_wake}
\end{figure}

\begin{figure} \centering
	\centerline{\includegraphics[scale=0.6]{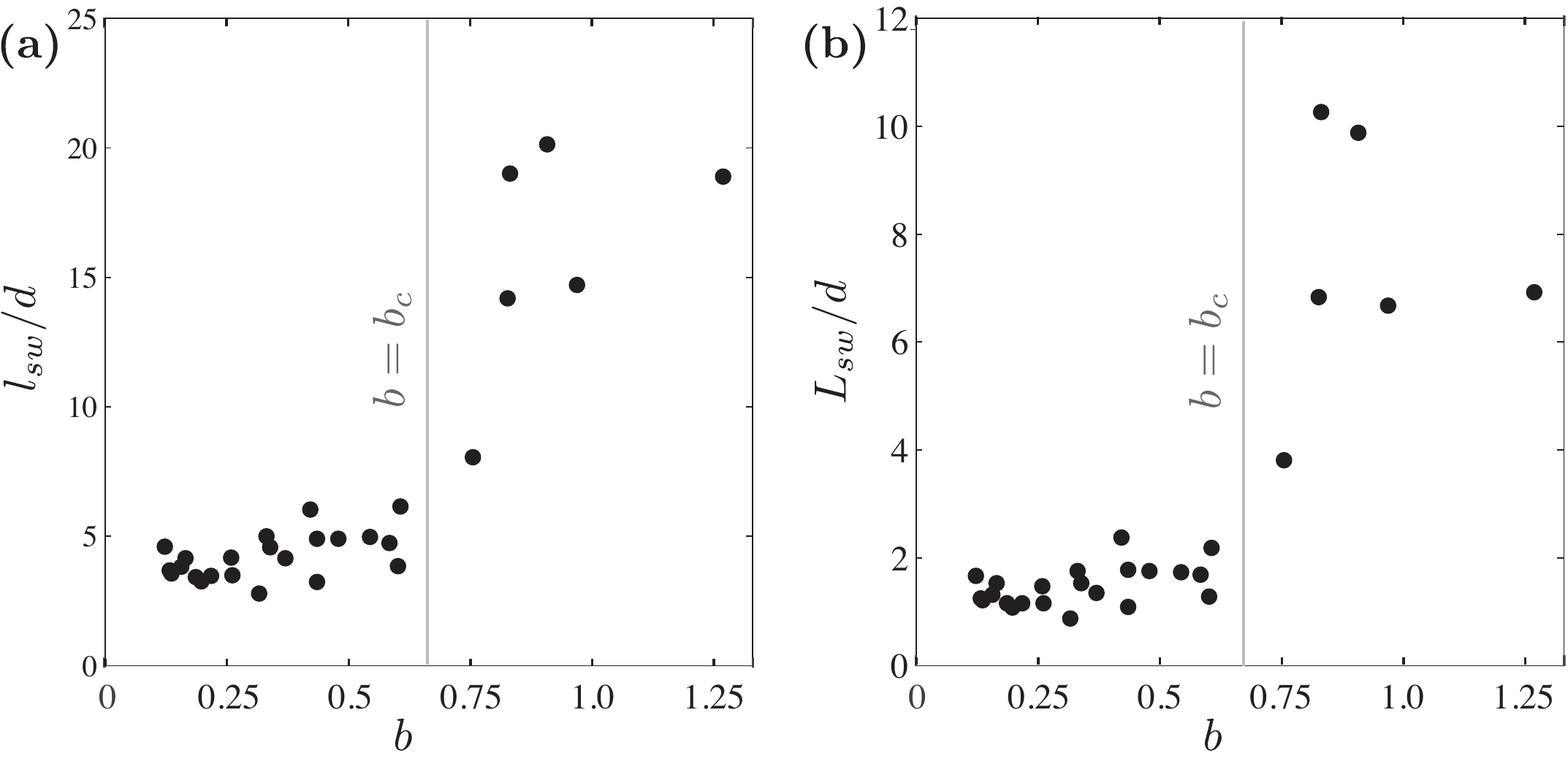}}
	\caption{Characteristics of the secondary wake. (a)~Characteristic length. (b)~Characteristic decay length.}
	\label{fig:cara_sec_wake}
\end{figure}

We now focus on the evolution of the characteristic lengths with the bubblance parameter~(figures \ref{fig:cara_near_wake} \& \ref{fig:cara_sec_wake}). Concerning the primary wake~(PW), the characteristic length $l_{pw}$ increases roughly linearly with $b$, while the characterisitic decay length $L_{pw}$ is almost independent of $b$ (figure \ref{fig:cara_near_wake} (a) \& (b)). The characteristic length $l_{pw}$ and the decay length $L_{pw}$ are thus controlled by two different mechanisms.
The characteristic length of the primary wake is controlled by the ratio between the two sources of turbulence and thus by $b$. On the contrary, since the decay length $L_{pw}$ does not depend on $b$, it means that the velocity deficit behind a bubble is not affected by the incident turbulence and is only controlled by the perturbation induced by the bubble itself. This is further clarified by the inset to figure~\ref{fig:wake}, which shows that the velocity deficit in the primary wake is not affected by the level of external turbulence. These findings thus extend the observations of \cite{legendre2006wake} for a single bubble in turbulence to the case of a dilute bubble swarm. Concerning the secondary wake, it presents different characteristics when $b$ increases (figure \ref{fig:cara_sec_wake} (a) \& (b)). In particular, the length of the secondary wake $l_{sw}$ is almost constant for $b<0.7$, after which it suddenly increases, reaching $\approx$ 20 $d$ for $b>0.8$. The decay length of the secondary wake $L_{sw}$ presents a similar behaviour. It is constant and close to the decay length of the primary wake for $b<0.7$, after which it suddenly increases. Consequently, the developement of the secondary wake is discontinuous with a threshold at around $b=0.7$, suggesting that the perturbation induced by the bubble should be strong compared to the external turbulence for the bubbles to develop a large secondary wake. However, the underlying physical mechanisms triggering the secondary remain unclear.

\subsection{Conditional pdf}

\begin{figure} \centering
	\centerline{\includegraphics[scale=0.6]{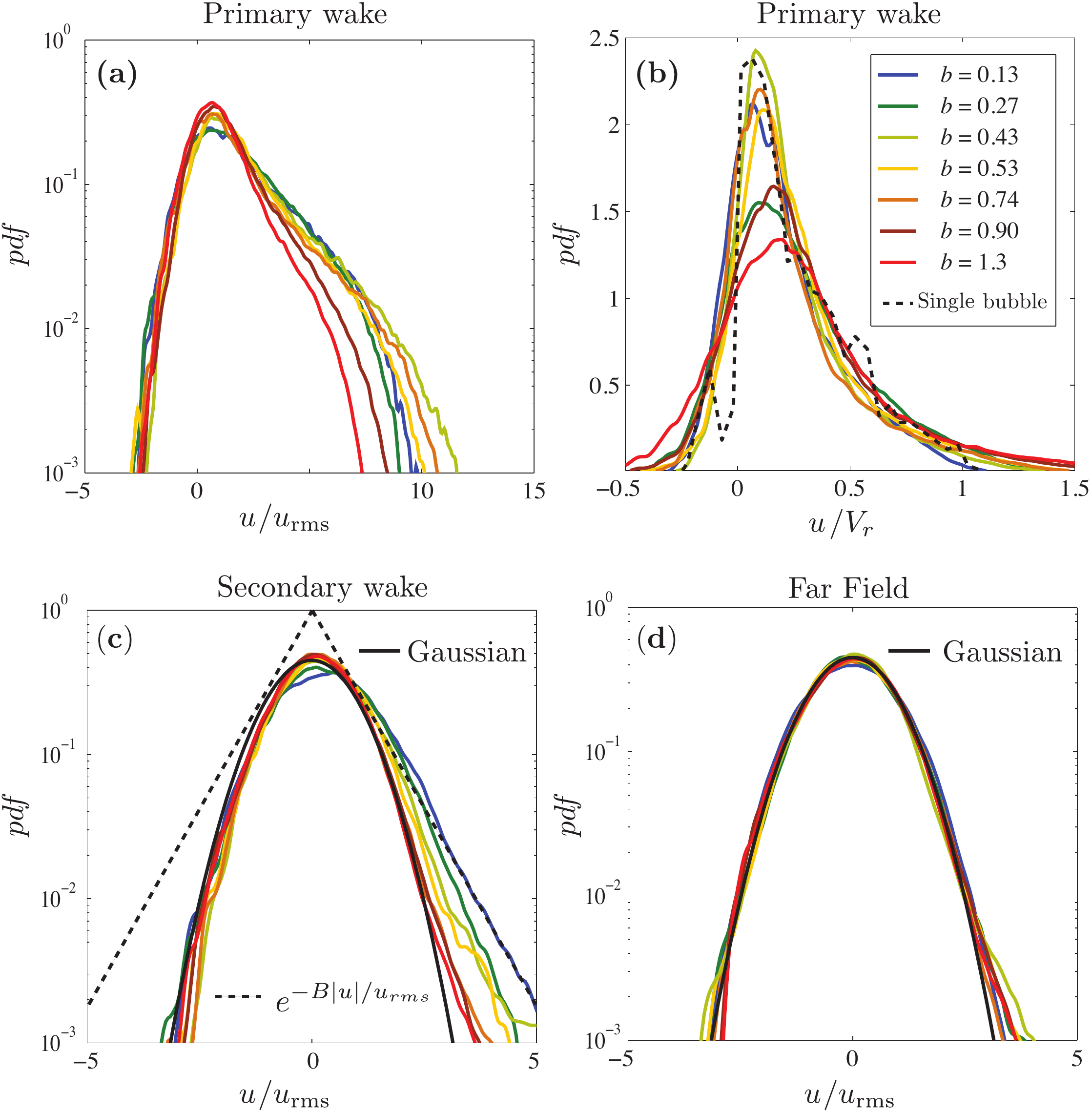}}
	\caption{PDF of the velocity fluctuations for different regions of the wake behind the bubble. (a) \& (b) Primary wake ~(PW), (c)~secondary wake~(SW), and (d)~far-field~(FF). \textcolor{black}{The dashed line in (c) represents an exponential distribution of the form $y =  e^{-B \mid u \mid/u_{rms}}$, where $B =1.25$, which fits well with the positive fluctuations of the $b =0.13$ case.}}
	\label{fig:CONDpdf_liq}
\end{figure}

We now carry out a specific study in the vicinity of the bubble. Three regions are distinguished: the primary wake, ($0<z<l_{pw}$), the secondary wake ($l_{pw}<z<l_{sw}$), and the far field ($l_{sw}<z$), referred to FF later. Conditional statistical properties of the velocity fluctuations, such as the standard deviation and the pdf, are calculated in each region by retaining the velocity fluctuations detected in those regions specifically. Figure~\ref{fig:CONDpdf_liq} (a), (c) and (d) presents the conditional pdf normalised by the corresponding standard deviation of the velocity fluctuations $u_{rms}$ for each region and different bubblance parameters. We observe that the shape of the conditional pdf strongly depends on the region under consideration. In fact, the conditional pdfs of the primary wake (PW) are asymmetric, presenting a strong skewness for positive values due to the strong perturbation induced by the bubbles. Remarkably, the pdf of the velocity fluctuations in the PW, once normalised by $V_r$, is similar to the one for a single bubble rising in a quiescent liquid~(see figure~\ref{fig:CONDpdf_liq} (b) \& \cite{risso2002velocity}). This means that the primary wake disturbance arises mainly from the bubble and is not much modified by the external turbulence. {\color{black} In contrast, the conditional pdfs in the far field are Gaussian, meaning that the external turbulence is dominant and that the wake-wake interactions are negligible in this area~(figure \ref{fig:CONDpdf_liq} (d)). Indeed, it is known from pseudo-turbulence, that the exponential tails of the pdf of the velocity fluctuations are the signature of the wake-wake interactions~\citep{risso2016physical}. Concerning the conditioned pdf of the secondary wake, they are asymmetric, since the tails are exponential for positive fluctuations and Gaussian for negative ones~(figure \ref{fig:CONDpdf_liq} (c)). In the present case, as the wake-wake interactions are negligeable, the secondary wake results thus mainly from the interaction of the wake with the external turbulence, generating exponential tail of the conditional pdf for the positive fluctuations only. In contrast to wake-wake interaction in pseudo-turbulence, the development of the interactions between the wake and the external turbulence is not an isotropic process in the vertical direction.}

\begin{figure} \centering
	\centerline{\includegraphics[scale=0.55]{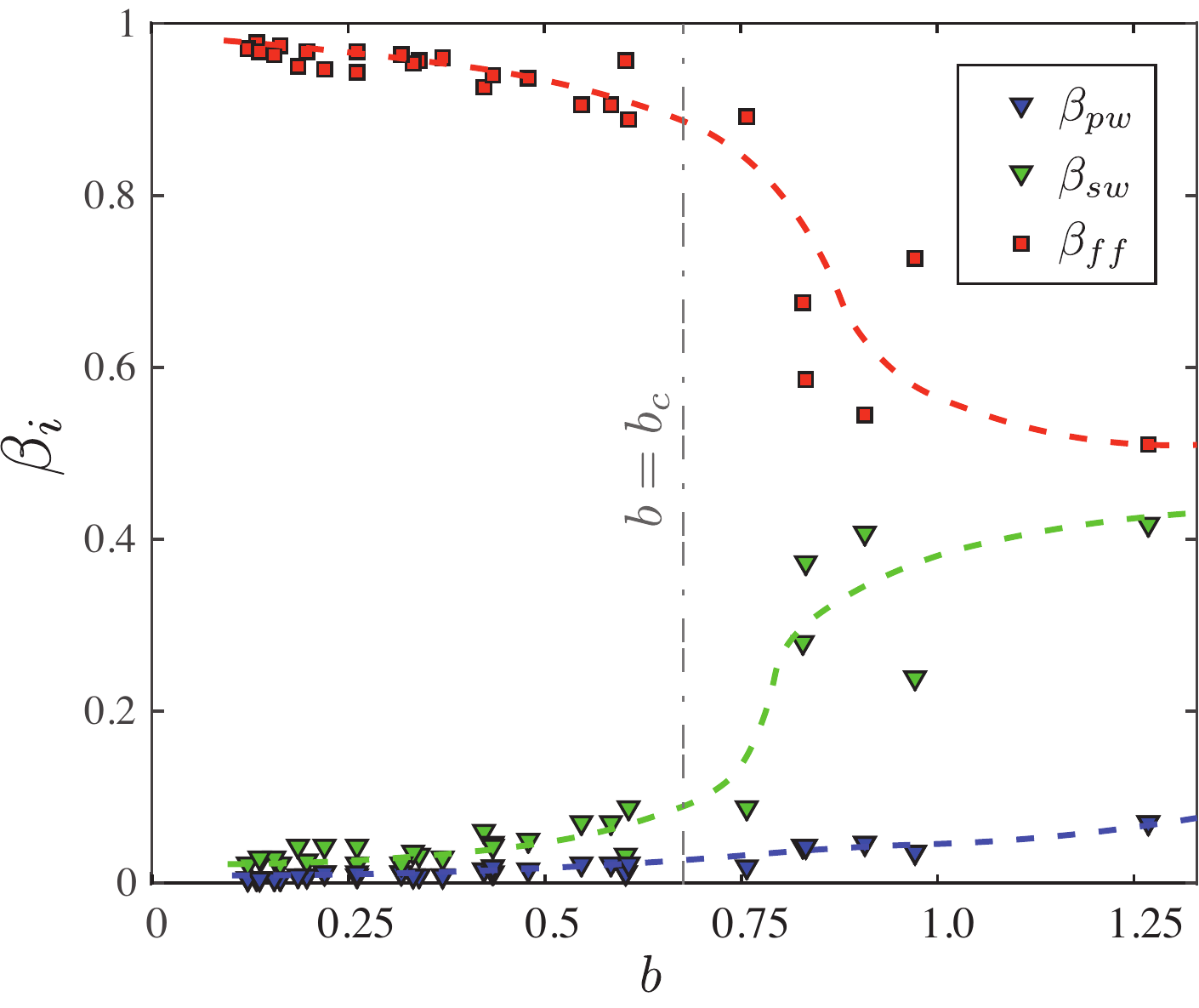}}
	\caption{Relative characteristic volume fraction ($\beta$) of the primary wake ($\beta_{pw}$), secondary wake ($\beta_{sw}$) and far field ($\beta_{ff}$). The dashed, dotted, and dash-dotted curves give our interpretation of the evolution of the PW, SW, and FF, respectively. The fractions sum to 1, i.e $\beta_{pw} + \beta_{sw} + \beta_{ff} = 1$.}
	\label{fig:volume_wake_d2b}
\end{figure}

\begin{figure} \centering
	\centerline{\includegraphics[scale=0.6]{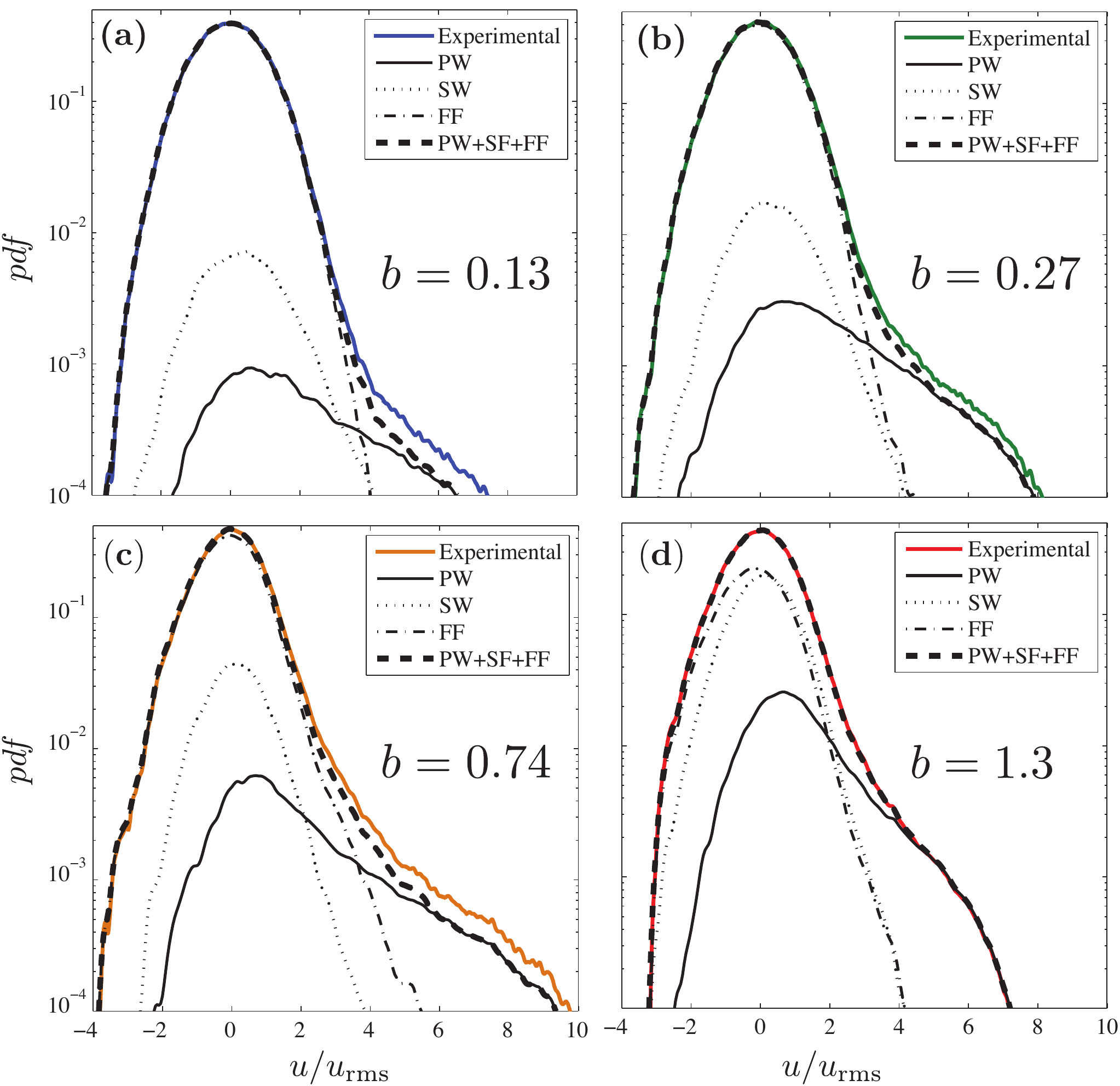}}
	\caption{Superposition of the conditional PDF weighted by their respective volumes for (a)~$b =0.13$, (b)~$b =0.27$, (c)~$b =0.74$, and (d)~$b =1.3$.}
	\label{fig:pdf_liq_SUPER}
\end{figure}

\section{Discussion}
\label{sec:discussion}

The pdf of the overall agitation in a turbulent bubbly flow results from the superposition of the conditional pdfs of the primary wake, the secondary wake and the far field, weighted by the volume of each region. Each conditional pdf presents different properties as shown in figure \ref{fig:CONDpdf_liq}. The question now is how to evalute the volume fraction of each region, especially since we only measured the length of the wakes in the vertical direction, and the lateral spread of the wake remains unknown. In order to estimate the volume of each region, the typical distance between two bubbles $d_{2b}$ is introduced. Assuming that the bubble are oblate ellipsoids with an aspect ratio $\chi$ and homogeneously distributed in space, the mean distance between two bubbles can be estimated as $d_{2b}= \frac{2 d}{3 \chi^{2/3} \alpha}$. The non-dimensional volume fraction $\beta$ of the primary wake (resp. secondary wake) may be written as $\beta_{pw}=\frac{l_{pw}}{d_{2b}}$ (resp. $\beta_{sw}=\frac{l_{sw}}{d_{2b}}$), and the volume fraction of the far field as $\beta_{ff}=1-(\beta_{sw}+\beta_{pw})$. Figure~\ref{fig:volume_wake_d2b} presents the volume fraction  of each region as a function of the bubblance parameter $b$. We can observe that even though the weight of the primary wake increases with $b$, it constitutes only a small fraction of the total liquid volume, less than 10$\%$, for the full range of $b$ investigated. In contrast, the secondary wake occupies almost 50$\%$ of the total volume when $b$ is larger than 0.7 The secondary wake thus plays a significant role at larger bubblance parameters.

Figure~\ref{fig:pdf_liq_SUPER} presents the superposition of each conditional pdf weighted by the respective fractions of each region for four contrasted bubblance parameters in the range $b =0.13-1.3$. We observe that superposing the three contributions allows us to get a realistic estimate of the total liquid agitation in the turbulent bubbly flow, even if the spreading of the wakes has not been taken into account. Thus, for low $b$, the agitation comes mainly from the far field, whereas it comes from both the far field and the secondary wake at larger values of $b$. However, irrespective of the value of $b$, the primary wake contributes mainly to the large positive fluctuations ranging from 4$u_{rms}$ to 8$u_{rms}$, causing the skewness of the pdf of the liquid fluctuations. This is similar to the observations of \cite{risso2002velocity} for pseudo-turbulence.

\section{Conclusion}
\label{sec:concl}

The hydrodynamic properties of a turbulent bubbly flow have been studied experimentally by varying both the level of the homogeneous and isotropic turbulence produced by the active grid~(from $u'_0=2.3$ cm/s to 5.5 cm/s) and the gas volume fraction~($\alpha$ from 0\% to 0.93$\%$). We found that the bubblance parameter is a suitable parameter to characterise such turbulent bubbly flows, as it compares the level of the turbulence induced by bubbles to that due to the external turbulence in absence of bubbles. The bubblance parameter is thus defined as $b = \frac{ V_r^2 \alpha}{u'^2_0}$ and varied in the range $0 - 1.3$.

We used the Constant Temperature Anemometry~(CTA) technique to look into the overall hydrodynamic properties {\color{black} in the vertical direction} of the turbulent bubbly flow. The spikes in the signal, which correspond to the gas-phase, were detected and removed, and we studied the evolution of the liquid phase velocity fluctuations with $b$. When normalised by the incident turbulent fluctuations, the velocity fluctuations collapse onto one curve, but show a non-monotonic evolution with $b$. For $b<0.27$, an attenuation of the turbulence is observed, while for $b > 0.27$, the standard deviation of the velocity fluctuations are enhanced. For the range $0.13 < b < 1.3$, we identified two regimes, with a transition appearing at a critical bubblance parameter, $b_c\approx 0.7$. For $b<b_c$, the normalised velocity fluctuations $u'/u_0'$ evolve as $\propto b^{0.4}$. In contrast, for $b>b_c$, the velocity fluctations show a steeper increase with b, i.e $\propto b^{1.3}$. The pdf of the velocity fluctuations, once normalised by the standard deviation, follows a Gaussian distribution for negative and more probable values, but with a slight skeweness for positive values. This skewness increases with $b$.

To reveal the origin of the two regimes and the skewness of the pdfs, we resorted to perform a conditional analysis on the turbulent bubbly flow. For this, we looked into the velocity fluctuations developing behind individual rising bubbles in the flow. This allowed us to distinguish three regions within the liquid phase: a primary wake, a secondary wake, and a far field. The primary wake, which is the region just behind the bubble, is mainly controlled by the bubble perturbation. As a consequence, the characteristic decay length of the primary wake is constant across all the bubblance parameters investigated. Furthermore, the pdf of the conditional velocity fluctuations of the primary wake normalised by the relative bubble velocity is similar to that of a single bubble rising in quiescent liquid. In contrast to this, the conditional pdf in the far field presents a Gaussian behaviour, meaning that it is mainly controlled by the incident turbulence. \textcolor{black}{Concerning the secondary wake, it results mainly from the interaction of the primary wake with the surrounding turbulence. The conditional pdf of the velocity fluctuations provides an illustration of this interaction, as they have nearly Gaussian tails for the negative fluctuations and exponential tails for the positive ones. The anisotropy of the conditional pdf is probably due to the presence of homogenous background turbulence. The properties of the secondary wake, such as its characteristic length and characteristic decay length present a sharp transition at the critical bubblance parameter $b_c \approx 0.7$. For $b < b_c$, the length of the secondary wake is $\approx 5 d$, and the remainder of the distance to the next bubble is occupied by the far-field, which has Gaussian fluctuations. This results in the weaker dependence of the overall velocity fluctuations on $b$, corresponding to the $u_{\text{rms}} \propto b^{0.4}$ regime. For $b>b_c$, the secondary wake is more developed and its length suddenly increases, reaching almost 20$d$. This means that the contribution of secondary wakes starts to become important for $b >b_c$, and therefore we see a stronger $b$ dependence in this regime~($u_{\text{rms}} \propto b^{1.3}$).}

The conditional pdfs also provide important clues about the origin of the skewness of the pdf. It turns out that  the skewness is due to the primary wake region behind the bubbles, since most of the fluctuations in the primary wake lie in the range of 4 to 8~$u_{rms}$. Furthermore, we observed a stronger skewness with increasing $b$, mainly due to the increase of the length of the primary wake with $b$. While the skewness of the pdf is the signature of the primary wake, the Gaussian shape of the pdf for negative fluctuations results from the far field and the secondary wakes fluctuations.

Another quantity that can help to disentangle the effect of the incident turbulence from the one induced by bubbles is the spectrum of the velocity fluctuations since it provides some crucial information on the time scales at which the incident turbulence and the one induced by bubbles act. In fact, for the present experimental conditions, the lower frequencies range: $1/T_L$ to $1/\tau_\eta$, the spectrum shows a $-5/3$ scaling, which is the classical evolution for homogeneous and isotropic turbulence. For frequencies larger than $V_r/\lambda$, where $\lambda=d/C_d$, the spectrum presents a $-3$ scaling, which is the signature of the turbulence induced by bubbles. Normalising the spectra with the frequency of the collective wake-instability of the bubble swarm~($f_{c} \approx 0.14 V_r/d$) makes the two ranges more clearly distinguishable for all the bubblance parameters studied here. \\

The present approach, which is based on the decomposition of the overall fluctuations into three contributions, has allowed us to understand the physical origin of dependences of the velocity fluctuation and the skewness of the pdf on $b$. In this regard, the sudden growth in the length and the decay length of the secondary wake at $b \approx 0.7$ is crucial to interpret our observations. As we have found here, the development of the secondary wake significantly enhances the overall turbulence in the liquid phase. Therefore, a better understanding of the physical mechanisms which trigger the secondary wake is required. This could be a key parameter for industrial spargers and bubble-column reactors, where more efficient mixing is desirable. In future work, we intend to link the liquid agitation behaviour observed here to the mixing of a passive scalar in turbulent bubbly flows.\\
	
	We thank Leen van Wijngaarden and Vivek Prakash for useful discussions, and Gert-Wim Bruggert and Martin Bos for technical support. This work is part of the industrial partnership programme of the Foundation for Fundamental
	Research on Matter (FOM). The authors also acknowledge the Netherlands Center for Multiscale Catalytic Energy Conversion~(MCEC), STW foundation, European High-performance Infrastructures in Turbulence~(EUHIT), and COST action MP1305 for financial support. Chao~Sun acknowledges the financial support from Natural Science Foundation of China under Grant No. 11672156.

\bibliographystyle{jfm}
\bibliography{mybib_huge.bib}

\end{document}